\documentclass[
reprint,
 superscriptaddress,
 amsmath,amssymb,
 aps,
prb,
]{revtex4-2}
\usepackage{tabularx}
\usepackage[colorlinks,bookmarks=false,citecolor=darkblue,linkcolor=red,urlcolor=blue]{hyperref}
\usepackage{graphicx}
\usepackage{dcolumn}
\usepackage{bm}
\usepackage{multirow}
\usepackage{booktabs}
\usepackage[dvipsnames]{xcolor}
\usepackage{upgreek}
\usepackage{boldline}
\usepackage{xcolor}

\definecolor{darkblue}{rgb}{0,0.02,0.45}

\newcommand{\fmo}{Fe$_{2}$Mo$_3$O$_8${}}

\newcommand{\cmo}{Co$_{2}$Mo$_3$O$_8${}}
\newcommand{\TN}{\ensuremath{T_{\mathrm{N}}}}

\newcommand{\Epara}{$\mathbf{E}^{\omega}\parallel c$}

\newcommand{\beginsupplement}{
    	\setcounter{table}{0}
    	\renewcommand{\thetable}{S\Roman{table}}
    	\setcounter{figure}{0}
    	\renewcommand{\thefigure}{S\arabic{figure}}
    	\setcounter{equation}{0}
    	\renewcommand{\theequation}{S\arabic{equation}}
        \setcounter{page}{1}
        \setcounter{section}{0}
        \begin{center}
            \textbf{\large Supplemental Materials: Optical phonons as a testing ground for spin group symmetries}
        \end{center}
    }

\usepackage[warn]{textcomp}
\usepackage{mathtools}

\DeclarePairedDelimiterX\braket[2]{\langle}{\rangle}{#1 \delimsize\vert #2}

\usepackage[version=4]{mhchem}
\usepackage{mathptmx}

\usepackage{amsmath}

\begin{document}

\title{Optical phonons as a testing ground for spin group symmetries}

\author{F.~Schilberth}
\thanks{These authors contributed equally}
\affiliation{Experimental Physics V, Center for Electronic
Correlations and Magnetism, Institute for Physics, University of Augsburg, D-86159 Augsburg, Germany}
\email{felix.schilberth@physik.uni-augsburg.de}
\affiliation{Department of Physics, Institute of Physics, Budapest University of Technology and Economics, M\H{u}egyetem rkp. 3., H-1111 Budapest, Hungary}
\author{M. Kond\'{a}kor}
\thanks{These authors contributed equally}
\affiliation{Department of Theoretical Physics, Institute of Physics, Budapest University of Technology and Economics, M\H{u}egyetem rkp. 3., H-1111 Budapest, Hungary}
\affiliation{Institute for Solid State Physics and Optics, HUN-REN Wigner Research Centre for Physics, H-1525 Budapest, P.O.B. 49, Hungary}
\author{D.~Ukolov}
\affiliation{Institute of Condensed Matter Physics, TU Braunschweig, Mendelssohnstr. 3, 38106 Braunschweig, Germany}
\author{A. Pawbake}
\affiliation{LNCMI, UPR 3228, CNRS, EMFL,
Université Grenoble Alpes, 38000 Grenoble, France}
\author{K.~Vasin}
\affiliation{Experimental Physics V, Center for Electronic
Correlations and Magnetism, Institute for Physics, University of Augsburg, D-86159 Augsburg, Germany}
\affiliation{Institute for Physics, Kazan (Volga region) Federal University, 420008 Kazan, Russia}
\author{O.~Ercem}
\affiliation{Experimental Physics V, Center for Electronic
Correlations and Magnetism, Institute for Physics, University of Augsburg, D-86159 Augsburg, Germany}
\author{L.~Prodan}
\affiliation{Experimental Physics V, Center for Electronic
Correlations and Magnetism, Institute for Physics, University of Augsburg, D-86159 Augsburg, Germany}
\author{V.~Tsurkan}
\affiliation{Experimental Physics V, Center for Electronic
Correlations and Magnetism, Institute for Physics, University of Augsburg, D-86159 Augsburg, Germany}
\affiliation{Institute of Applied Physics, Moldova State University, str. Academiei 5, MD2028, Chisinau, Republic of Moldova}
\author{A.A.~Tsirlin}
\affiliation{Felix Bloch Institute for Solid-State Physics, Leipzig University, 04103 Leipzig, Germany}
\author{C. Faugeras}
\affiliation{LNCMI, UPR 3228, CNRS, EMFL,
Université Grenoble Alpes, 38000 Grenoble, France}
\author{P.~Lemmens}
\affiliation{Institute of Condensed Matter Physics, TU Braunschweig, Mendelssohnstr. 3, 38106 Braunschweig, Germany}
\affiliation{Laboratory for Emergent Nanometrology (Lena), TU Braunschweig, Langer Kamp 6, 38106 Braunschweig, Germany}
\author{K. Penc}
\affiliation{Institute for Solid State Physics and Optics, HUN-REN Wigner Research Centre for Physics, H-1525 Budapest, P.O.B. 49, Hungary}
\author{I.~K\'ezsm\'arki}
\affiliation{Experimental Physics V, Center for Electronic
Correlations and Magnetism, Institute for Physics, University of Augsburg, D-86159 Augsburg, Germany}
\author{S.~Bord\'{a}cs}
\affiliation{Experimental Physics V, Center for Electronic
Correlations and Magnetism, Institute for Physics, University of Augsburg, D-86159 Augsburg, Germany}
\affiliation{Department of Physics, Institute of Physics, Budapest University of Technology and Economics, M\H{u}egyetem rkp. 3., H-1111 Budapest, Hungary}
\affiliation{HUN-REN–BME Condensed Matter Physics Research Group, Budapest University of Technology and Economics, M\H{u}egyetem rkp. 3., H-1111 Budapest, Hungary}
\author{J.~Deisenhofer}
\email{joachim.deisenhofer@physik.uni-augsburg.de}
\affiliation{Experimental Physics V, Center for Electronic
Correlations and Magnetism, Institute for Physics, University of Augsburg, D-86159 Augsburg, Germany}

\date{\today}

\begin{abstract}
Lattice vibrations are highly sensitive to crystal symmetries and their changes across phase transitions. The latter can modify irreducible (co)representations and corresponding infrared and Raman selection rules of phonons. This concept is established for relativistic magnetic point groups, simultaneously transforming spatial and spin coordinates. However, in altermagnets described by non-relativistic spin groups with disjunct symmetry operations for both vector spaces, the phonon selection rules have remained unexplored.
Here, we present a detailed study of the infrared- and Raman-active modes in the collinear antiferromagnet and altermagnet candidate \cmo{}. Comparing to \textit{ab initio} calculations accurately capturing the eigenfrequencies, we identify all expected phonon modes at room temperature and deduce their selection rules using both symmetry approaches. Importantly, we observe the change of selection rules upon antiferromagnetic ordering, agreeing with the relativistic symmetry approach, while the spin group formalism predicts no changes. Therefore, optical phonons can reveal the appropriate symmetry treatment.
\end{abstract}

\maketitle

\section{Introduction}

The realm of compensated collinear antiferromagnets has been shaken by the introduction of the concept of altermagnetism \cite{Smejkal:2022} and many materials, which previously were regarded as benchmark ``Néel" antiferromagnets such as the insulating rutiles MnF$_2$ and CoF$_2$ are now reconsidered for signatures of altermagnetism \cite{Bhowal:2024,Morano:2024, Yu:2024,Dubrovin:2024}, in particular for non-relativistic spin splittings of electronic bands or magnons along general directions in the Brillouin zone. In this respect, MnTe appears to be among the most prominent realizations of such splittings \cite{Lee:2024,Krempasky:2024,Liu:2024,Jost:2025}, while the altermagnetism in RuO$_2$ remains controversial \cite{Fedchenko:2024,Wenzel:2025, Liu2024}.

In terms of symmetry, altermagnets and the breaking of Kramers degeneracy at a general point in the Brillouin zone are described by using the spin group concept introduced in the 1960s and classified by Litvin and Opechowski \cite{Litvin:1974,Litvin1977}, which separates spatial symmetry operations and symmetry elements in spin space corresponding to the non-relativistic limit with zero spin-orbit coupling.

This limit is in clear contrast to the concept of the Shubnikov magnetic space and point groups. Those are commonly used to determine the transformation properties of observables by applying the Neumann principle and selection rules for excitations. The magnetic point group elements simultaneously leave the structural and the spin configuration invariant, i.e. a relativistic setting with non-zero spin-orbit coupling is imposed \cite{Bradley:2009}.

We want to compare these two symmetry approaches by investigating experimentally and theoretically the optical phonons and their selection rules in \cmo{}, which  belongs to the family of polar molybdenum oxides $A_{2}$Mo$_3$O$_8$  ($A$ = Mn, Fe, Co, Ni, Zn). These polar materials exhibit different magnetically ordered ground states, which can be tuned by external magnetic fields or doping \cite{Kurumaji:2015,Wang:2015,Kurumaji:2017,Tang:2019,Csizi:2020,Tang:2021,Tang:2022,Reschke:2022,Prodan:2022,Ghara:2023,Szaller:2025}. For \fmo{}, low-lying chiral phonons and magnon-polariton excitations in the THz frequency range have been reported \cite{Kurumaji:2017a,Wu:2023,Bao:2023,Vasin:2024}, which exhibit non-reciprocal directional dichroims \cite{Reschke:2022,Vasin:2024}. Only recently, it was recognized that the collinear antiferromagnetic phases of \cmo{} and \fmo{} fulfill the necessary criteria for altermagnets \cite{Cheong2024} and even ``altermagnetoelectric" effects have been predicted to occur \cite{Smejkal:2024}.

In the paramagnetic regime, \cmo{} and the other family members $A_{2}$Mo$_3$O$_8$  ($A$ = Mn, Fe, Co, Ni, Zn) crystallize in the non-symmorphic polar hexagonal space group $P6_3mc$ (\#186), featuring a built-in polarization along the $c$-axis. The unit cell is shown in Fig. \ref{fig:structure}(a). The $A^{2+}$ ions are responsible for magnetism, with half of them occupying the corner-sharing tetrahedral (A) and the other half the octahedral (B) sites.
The Mo ions build non-magnetic trimers \cite{Varret:1972,Cotton:1964,Wang:2015}.
Upon cooling, \cmo{} undergoes a collinear antiferromagnetic order at $\TN{}=39$~K \cite{Tang:2019,Reschke:2022,Tang:2022,Prodan:2022, Szaller:2025}, the sister compound \fmo{} at $\TN{}=60$~K \cite{Kurumaji:2015,Wang:2015}, with all spins aligned parallel to the $c$-axis as illustrated in Fig.~\ref{fig:structure}. No structural symmetry changes were observed upon entering the AFM phase for both compounds \cite{Tang:2019,Reschke:2020}, making them ideal candidates to study symmetry changes due to the collinear antiferromagnetic ordering.

First, we analyze the symmetry  properties and  selection rules in \cmo{}  for Raman- and infrared (IR)-active phonons in terms of the relativistic magnetic point group and the non-relativistic spin group approach.
Then, we discuss the optical modes observed by IR- and Raman spectroscopy and identify the phonons by comparison of the eigenfrequencies with \textit{ab initio} calculations and conclude on the applicability of the two symmetry concepts for \cmo.

\section{Results}

\subsection{Symmetries of the paramagnetic and the collinear magnetic state of  \cmo{}}

\begin{figure*}[t]
\includegraphics[width = 0.9\textwidth]{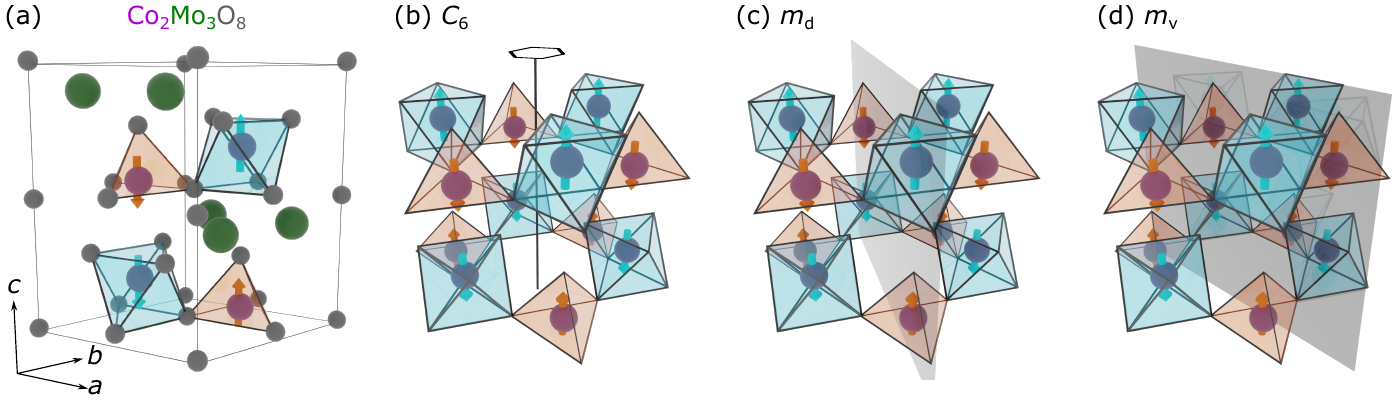}
\caption{\label{fig:structure} Crystal structure and magnetic ordering of \cmo: (a) Crystallographic and magnetic unit cell with two formula units illustrating the collinear $q=0$ antiferromagnetic spin order. (b)-(d) Symmetry elements generating the crystallographic point group $6mm$: (b) the six-fold rotation axis $C_6$, (c) the diagonal mirror plane $m_d=\{m_{100},m_{010},m_{1\bar{1}0}\}$, and (d) the vertical mirror plane $m_v=\{m_{110},m_{120},m_{2\bar{1}0}\}$. When combined with a half-translation along the $c$ axis, the $C_6$ forms a screw axis, and the $m_v$ becomes a glide plane. They are symmetry elements of the space group $P6_3mc$ (No.186).}
\end{figure*}

The crystallographic point group in both the paramagnetic and magnetically ordered phases is $G=6mm$ ($C_{6v}$), which contains twelve symmetry operations grouped into six conjugacy classes:
\begin{align}
   \mathbf{G}&=\{E\}+ \{C^+_6,C^-_6\}+\{C^+_3,C^-_3\}+ \{C_2\} \nonumber\\
   &\phantom{=}+\{m_{100},m_{010},m_{1\bar{1}0}\}
   %\nonumber\\&\phantom{=}
   + \{m_{110},m_{120},m_{2\bar{1}0}\}\nonumber\\
   &= \{E\}+ \{C_6\}+\{C_3\}+ \{C_2\}+ \{m_\mathrm{d}\}+ \{m_\mathrm{v}\}
\end{align}
Fig.~\ref{fig:structure}(b)-(d) shows the symmetry generators of the $6mm$ point group. Note that the $C_6$ sixfold and the $C_2$ twofold rotations, as well as the  $m_v$ mirror planes are non-symmorphic: each must be accompanied by a half-translation $(c/2)$ along the $z$-axis, transforming them into screw-axis rotations and glide-plane reflections in the non-symmorphic space group $P6_3mc$.

Superimposing the collinear spin configuration of the magnetically ordered state onto the corresponding $A$- and $B$-sites occupied by the magnetic ions, one can follow two routes.
(i) The relativistic route, where finite spin–orbit interactions lead to a description using magnetic space and point groups. For \cmo{}, this is determined by the halving subgroup $\mathbf{H}=3m$ of $\mathbf{G}$, yielding the magnetic point group $\mathbf{M}=6^\prime m^\prime m$. Due to the Lorentz-invariant relativistic framework, the simultaneous transformation of spatial coordinates and spin degrees of freedom naturally arises from the spin-orbit coupling. Alternatively, we can take the non-relativistic route (ii), where we neglect spin–orbit effects and describe the order in terms of spin groups, as is done for altermagnets. If we infer the collinear spin arrangement on the two sublattices which in Litvin's notation is denoted by a superscript $\bar{1}$ for crystallographic symmetries connecting the two sublattices \cite{Litvin1977}, we obtain the spin group $\mathbf{G}_s = \phantom{i}^{\bar{1}}6^{1}m^{\bar{1}}m$ (The classification for both approaches is detailed in Supplement \ref{app:grouptheory}).

Remarkably, the group-theoretical condition for altermagnetism can be expressed using irreducible representations of the crystallographic point group $\mathbf{G}$ and the primary antiferromagnetic order parameter given by the N\'eel vector $\mathbf{L}$: If $\mathbf{L}$
belongs to a one-dimensional real representation of $\mathbf{G}$, which remains invariant under all operations preserving the momentum $\mathbf{k}$, then the antiferromagnetic order is compatible with altermagnetism \cite{McClarty:2024,Schiff:2024}. In \cmo{}, the magnetic ordering exhibits a complex structure due to the presence of spins located on both tetrahedral (A sites) and octahedral (B sites) sublattices, with Néel order established independently on each sublattice. The Néel vector can be defined as $\mathbf{L}=\mathbf{L}_{A}+\mathbf{L}_{B}=\frac{1}{2}(\mathbf{M}_{1}-\mathbf{M}_{2})$, where $\mathbf{L}_{X}=\frac{1}{2}(\mathbf{M}_{1 X}-\mathbf{M}_{2 X})$ denotes the N\'eel vectors on $X=A,B$ sites, and subscripts 1 and 2 denote the magnetization of the magnetic sublatices (specifically, subscript 1 corresponds to the up-spin magnetizations and 2 to the down-spin magnetizations) \cite{Ghara:2023}. Depending on the choice of the isomorphism between $\mathbf{G}$ and the abstract group $6mm$, the Néel vector transforms according to either the irreducible representation $B_1$ or $B_2$ of $6mm$ (see Tab.~\ref{tab:6mm} in Supplement \ref{app:grouptheory}). In both scenarios, $\mathbf{L}$ changes sign only under symmetry operations that do not map $\mathbf{k}$ to $-\mathbf{k}$. This implies that inversion and time-reversal do not simultaneously remain symmetries, hence \cmo{} meets the group-theoretical requirement for altermagnetism.

The key difference between the classifications (i) and (ii) for our study is how the magnetic groups modify the corepresentations and selection rules for optically active phonons, which we detail in the following.

\subsection{Irreducible (co)representations and selection rules for IR- and Raman-active phonons}
\label{sec:phonons}

In this section, we discuss phonon selection rules in \cmo{} by considering the symmetry of the paramagnetic and magnetically ordered phases. In particular, we will outline the difference in phonon selection rules arising from the use of the non-relativistic spin point group for altermagnets in comparison with the magnetic point group for spin-orbit coupled systems.
While unitary representation theory adequately describes crystallographic point groups, magnetic point groups require corepresentation theory to include the antiunitary operations. Corepresentations play an analogous role for magnetic groups to that of unitary representations for nonmagnetic groups and are conventionally denoted as $D\Gamma$ \cite{Bradley:2009}. The symmetry requirements for IR- and Raman-active modes are summarised in Supplement \ref{app:coreps}.

Explicitly, the number and symmetry of the allowed optical phonon modes for paramagnetic \cmo{}  with two formula units in the primitive unit cell \cite{Tang:2019} are given by
\begin{align}\label{Eq:irreps}
\Gamma &= 9A_1(z;xx,yy,zz)   + 12E_1(x,y;xz,yz) &&\text{(IR + Raman)}\nonumber \\ 
&+  13E_2 (x^2-y^2,xy) &&\text{(Raman)}\nonumber \\  
&+  A_1 +E_1 &&\text{(acoustic)}\nonumber \\ 
&+ 3A_2 + 10B_1 +3B_2  &&\text{(silent),}     
\end{align}
where the IR activity corresponds to the linear coordinate functions of a polar vector and the quadratic coordinate functions to the Raman activity of the polar second-rank Raman tensor. Hence, in the IR experiments, nine $A_1$-modes should be observable for \Epara{}  and in the Raman channel these modes should appear exclusively in the configurations $x(zz)\bar{x}$ or $y(zz)\bar{y}$. The 12  $E_1$-modes are IR-active for $\mathbf{E}^{\omega}\parallel a$  and can be identified in the Raman configurations $y(xz)\bar{y}$ or $x(yz)\bar{x}$. The 13 $ E_2$ Modes are only Raman active and can be singled out in the Raman channel $z(xy)\bar{z}$. Hence, the standard experimental procedure to identify the IR and Raman-active phonon modes is to measure IR reflectivity in the two polarization configurations and Raman scattering in the three different Raman configurations, as it has been performed in previous IR and Raman studies of the isostructural compound \fmo{} \cite{Reschke:2020,Stanislavchuk:2020}. In addition, we perform Raman measurements in the $z(xx)\bar{z}$ configuration, which, in the paramagnetic phase, allows the observation of the $A_1$ and $E_2$ modes.

Upon magnetic ordering, the symmetry is lowered from the paramagnetic symmetry group to the magnetic space group $P6^\prime_3 m^\prime c$ with the magnetic point group $6^\prime m^\prime m$, but the primitive unit cell and, consequently, the total number of normal modes remain unchanged. In analogy to the paramagnetic case, the selection rules have to be determined now using the irreducible corepresentations of the antiferromagnetic state given by
\begin{align}\label{Eq:coreps}
D\Gamma &= 19DA_1(z;xx,yy,zz) &&\text{(IR+Raman)} \nonumber\\
&+ 25DE(x,y;xx,yy,xy,xz,yz) &&\text{(IR+Raman)}\nonumber \\ 
&+  DA_1 +DE &&\text{(acoustic)}\nonumber \\ 
&+ 6DA_2 &&\text{(silent).}     
\end{align}
The corresponding symmetry adapted Raman tensors for the 58 black-and white magnetic point groups are given by Cracknell \cite{Cracknell:1969} and summarised for \cmo{} in Tab.~\ref{tab:Ramantensors} in Supplement~\ref{app:grouptheory}. The $A_1$ and $B_1$ modes will be contained in the corepresentation  $DA_1$, the $A_2$ and $B_2$ modes form the class $DA_2$, and $E_1$ and $E_2$ will form the $DE$ class. Consequently, upon cooling into the AFM phase one may expect many more modes to become visible both in IR- and Raman measurements.

As outlined above, \cmo{} is a candidate for altermagnetism. Therefore, we may ask whether we can obtain different selection rules if we consider spin-group symmetries instead of the usual magnetic point groups, and if they are realized in the material.
As shown in Ref.~\onlinecite{Schiff:2025}, the spin point group $^{\bar{1}}6^{\bar{1}}m^{1}m$ introduces new irreducible corepresentations compared to the magnetic point group, (both tabulated in Table~\ref{tab:char-spingr6mm} in Supplement~\ref{app:grouptheory}), possibly giving rise to new or additional selection rules. 
To derive these for IR and Raman active phonons, we begin by comparing the character tables of the crystallographic point group $6mm$ (Table~\ref{tab:6mm}), its magnetic counterpart $6'm'm$ (Table~\ref{tab:3m}), and full spin group $^{\bar{1}}6^{\bar{1}}m^{1}m$ (Table~\ref{tab:char-spingr6mm}). Introducing magnetic order halves the number of unitary operations and promotes the remaining irreps of $6mm$  into co‑irreps of the unitary subgroup $3m$. Antiunitary elements follow the co-representation relation
\begin{equation}
    D\Gamma(au)=D\Gamma(a)(D\Gamma(u))^{\star} \;,
\end{equation}
instead of the usual product rule $\Gamma(uv)=\Gamma(u) \Gamma(v)$ with 
 $u,v$ denoting unitary and $a$ antiunitary symmetry elements \cite{Bradley:2009}. 
 However, as soon as spin and real space components decouple and new unitary transformations arise, one can associate each unitary operation of the spin group $\left[s||g\right]$ in $^{\bar{1}}6^{\bar{1}}m^{1}m$ with $g$ in $6mm$. Here, $s$ is the spin space operation and $g$ the real space operation of the spin group symmetry. As a result, all irreps of $6mm$ can be extended to co-irreps in $^{\bar{1}}6^{\bar{1}}m^{1}m$ -- these are given by $D\Gamma_1- D\Gamma_6$ in Table~\ref{tab:char-spingr6mm}. 

Because the electric‐dipole polarization $\mathbf{p}$,  the electric field of light $\mathbf{E}^\omega$, and the symmetric Raman tensor $\hat{R}$ transform purely under spatial operations as shown in Supplement \ref{app:light_transform}, they correspond exclusively to the six co‑irreps $D\Gamma_1-D\Gamma_6$ that can be directly associated to the irreducible representations of the crystallographic point group (or the co-irreps of the paramagnetic group). Consequently, the spin‐group analysis reproduces exactly the crystallographic selection rules of $6mm$ seen in the paramagnetic phase.

At the microscopic level, phonons sense magnetic order via spin–phonon interactions arising from slight distortions of the ligand environment around each magnetic ion. These terms enter the unperturbed Hamiltonian and, if they respect the full spin‑group symmetry, will enforce the corresponding spin‑group selection rules with respect to the electric‑dipole perturbation $\mathbf{E} \cdot \mathbf{p}$. Conversely, if the spin–phonon coupling requires the coupled spin and real-space transformations, then the allowed IR and Raman transitions revert to those prescribed by the magnetic point group symmetry alone.

In conclusion, our analysis demonstrates that the selection rules of the corresponding spin group for IR and Raman active phonons are identical to those determined by its crystallographic point group. As a result, such probes provide information about the importance of the spin-orbit coupling and an educated guess about the influence of relativistic effects on magnonic or electronic band structures. In the case of \cmo{} the crystallographic point group does not change upon magnetic ordering and hence, no changes are expected between the paramagnetic and antiferromagnetic phases. This result clearly contrasts with the predictions derived from the relativistic magnetic point group analysis presented in Eq.~(\ref{Eq:coreps}).

\subsection{Raman and Infrared active modes in \cmo}

\begin{table}[b]
\caption{\label{tab:zz}
Comparison of experimental IR and Raman active  excitation frequencies (in cm$^{-1}$) in \cmo{} measured for light polarization \Epara{} and in the Raman configuration $y(zz)\bar{y}$, respectively. The experimental eigenfrequencies measured at 295\,K and 10\,K are compared with theoretical \textit{ab initio} values for the nine expected $A_1$ modes.}
\begin{ruledtabular}
\begin{tabular}{cc|cc|c}
\multicolumn{2}{c|}{IR}         & \multicolumn{2}{c|}{Raman}   &  Mode \\
\multicolumn{2}{c|}{\Epara{}}   & \multicolumn{2}{c|}{$y(zz)\bar{y}$}  & assignment      \\
295\,K & 10\,K                  & 295\,K & 10\,K               & $A_{1}$(i) $i=1,\dots,9$  (calc.)  \\
\hline
 -   &  -   & 205 & 207 & 204 \\
264  & 263  & 264 & 264 & 271 \\
367  & 370  & 369 & 372 & 356 \\
446  & 448  & 446 & 449 & 453 \\
467  & 468  & 463 & 466 & 456 \\
565  & 569  & 560 & 564 & 564 \\
653  & 656  & 652 & 655 & 652 \\
731  & 733  & 729 & 732 & 723 \\
793  & 793  & 786 & 789 & 811 \\
\hline
 -   & 304  &  -  & 304 &  electronic  \\
\end{tabular}
\end{ruledtabular}
\end{table}

In this section we will compare the measured IR- and Raman spectra of \cmo{} obtained in the paramagnetic phase (at 295~K or 300~K) and in the antiferromagnetic phase (at 5~K or 10~K) with the calculated eigenfrequencies obtained by ab initio calculations, in order to identify newly activated modes of the magnetically ordered state. As the overall number of expected phonon modes remains constant across the magnetic transition, we will refer to the phonon modes using the nomenclature of the irreducible representations of the non-magnetic crystallographic point group for clarity reasons. However, one should always be aware that modes which belong to different irreducible representations in the paramagnetic state may belong to the same irreducible corepresentation in the magnetically ordered state, as discussed above. 

It is also important to note that we compare the Raman scattering results with the expected selection rules as given by Eqs.~(\ref{Eq:irreps}) and (\ref{Eq:coreps}) derived for the approximate first-order non-resonant Raman cross section for non-polar modes as discussed above. Strictly speaking, these Raman selection rules are only justified for the non-polar $E_2$-modes in the paramagnetic phase, but the main reported effects of the internal electric field originating from polar phonons in uniaxial crystals like \cmo{} is the lifting of degeneracies and shifts of such modes for scattering angles away from the backscattering configuration used in this study \cite{Shapiro:1972,Ovander:1979,Hayes:2004}. Hence, we do not expect to observe scattering effects of the polar nature of the phonons.

%\onecolumngrid

\begin{figure*}[t]
    \centering
    \includegraphics[width=\linewidth]{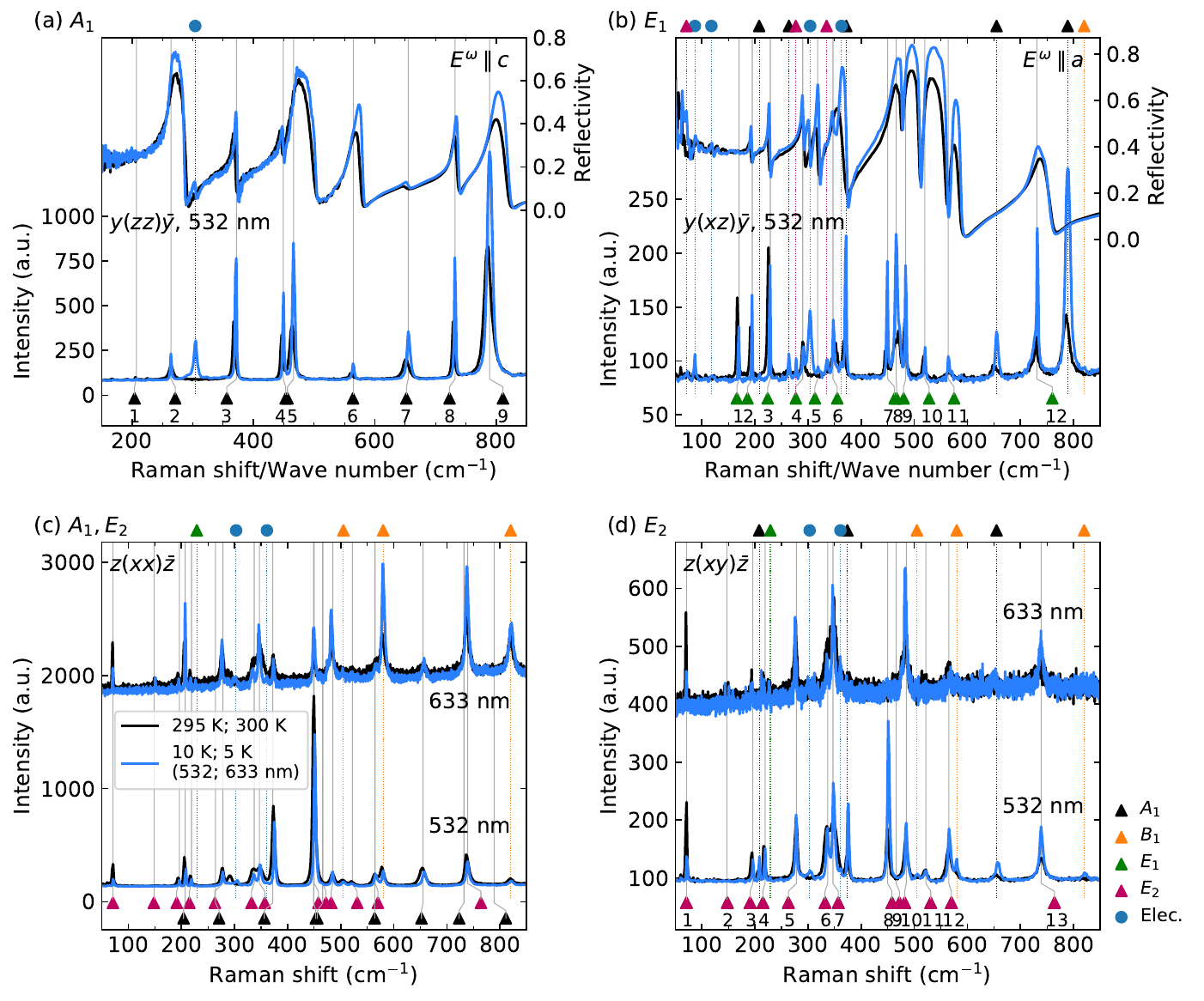}
    \caption{Comparison of Raman and  far-infrared reflectivity spectra at 295 K/300 K (black) and 10 K/5 K (blue) in (a) scattering configuration $y(zz)\bar{y}$ (532~nm) and $\mathbf{E}^\omega\parallel c$ corresponding to the irreducible $A_1$ representations, (b) in scattering configuration $y(xz)\bar{y}$ (532~nm) and $\mathbf{E}^\omega\parallel a$ corresponding to the irreducible $E_1$ representation. (c), (d) Comparison of Raman spectra at laser wavelengths of 532~nm and  633~nm in scattering configurations $z(xx)\bar{z}$ allowing for the observation of $E_2$ and $A_1$ modes, and $z(xy)\bar{z}$ corresponding to the irreducible $E_2$ representations of the paramagnetic phase, respectively. The calculated phonon eigenfrequencies expected for each configuration in the paramagnetic phase are indicated by upward triangles at the bottom of the individual panels. Symbols at the top of the panels indicate modes different from expected modes of the paramagnetic phase as described in the text.}
    \label{fig:comp_IR_532nm}
\end{figure*}

%\twocolumngrid

\begin{table}[t]
\caption{\label{tab:xz}
Comparison of experimental IR and Raman active  excitation frequencies (in cm$^{-1}$) in \cmo{} measured for light polarization $\mathbf{E}^\omega\parallel a$ and in the Raman configuration $y(xz)\bar{y}$, respectively. Mode assignment is made by comparison with calculated phonon eigenfrequencies for $E_1$ and $B_1$ modes and by comparing to experimental values from other Raman measurement configurations.}
\begin{ruledtabular}
\begin{tabular}{cc|cc|c}
\multicolumn{2}{c|}{IR}         & \multicolumn{2}{c|}{Raman}   & Mode   \\
\multicolumn{2}{c|}{$E^{\omega}\parallel a$}   & \multicolumn{2}{c|}{$y(xz)\bar{y}$}  &  assignment    \\
295\,K & 10\,K                  & 295\,K & 10\,K               & $E_{1}$(i) $i=1,\dots,12$ (calc.)\\
\hline
 -   &  -     &  167 &   170 &   166 \\
191  &  193    &  191 &   195 &   186 \\
225  &  227    &  225 &   229 &   224 \\
289  &  290    &  290 &   291 &   277 \\
315  & 317     &  314  &  318  &   313 \\
354  &  345    &  352 &   347 &   355 \\
 -   &  -      &  445 &   450 &   463 \\
465  & 463     &  470 &   466 &   467 \\
487  & 486     &  482 &   484 &   480 \\
527  &  521    &  517 &   520 &   528 \\
568  &  570    &  561 &   564 &   575 \\
737  &  728    &  730 &   731 &   760 \\
\hline
 -   &  -      & 205 &   208 & $A_1$(1) at 207 (exp.) \\
 -   &  -      & 264 &   264 & $A_1$(2) at 264 (exp.) \\
  -   &  -     & 367 &   372 & $A_1$(3) at 372 (exp.) \\
 -   &  -      & 650 &   655 & $A_1$(7) at 655 (exp.) \\
 -   &  -      & 786 &   789 & $A_1$(9) at 789 (exp.) \\
\hline
 -   &  -  &  -  &   72 & $E_2$(1) at 71 (exp.) \\
 -   &  87     &   -  &   87 & electronic \\
 -   & 118  &  -  &   118  & electronic \\
 -   &  -      & - &  277  & $E_2$(5) at 278 (exp.) \\
  -   &  301    &  -  &  304  & electronic \\
  -   &  -      & - &  335  & $E_2$(6) at 337 (exp.) \\
 -   &  361    &  -  &  363  & electronic \\
   -   &  -    &  -  &  820  & $B_1$(10) at 840 (calc.) \\
\end{tabular}
\end{ruledtabular}
\end{table}

\begin{table*}[htb]
\caption{\label{tab:xyresonant}
Experimental Raman excitation frequencies in \cmo{} (in cm$^{-1}$) measured in  $z(yx)\bar{z}$ and  $z(xx)\bar{z}$ configuration in the paramagnetic phase (at 295~K or 300~K) and at low temperatures in the magnetically orderd phase (at 5~K or 10~K) with different laser frequencies. Mode assignment is made by comparison with calculated phonon eigenfrequencies obtained from \textit{ab initio} calculations for the 13 expected $E_2$ modes and by comparing to experimental values from other measurements configurations.}
\begin{ruledtabular}
\begin{tabular}{cc|cc|cc|cc|c|cc|cc|cc|cc}
\multicolumn{8}{c|}{Raman: $z(xy)\bar{z}$} & Mode &\multicolumn{8}{c}{Raman: $z(xx)\bar{z}$} \\
\multicolumn{2}{c|}{632.8~nm} & \multicolumn{2}{c|}{532~nm} & \multicolumn{2}{c|}{514.3~nm} & \multicolumn{2}{c|}{473~nm} & assignment & \multicolumn{2}{c|}{632.8~nm} & \multicolumn{2}{c|}{532~nm} & \multicolumn{2}{c|}{514.3~nm} & \multicolumn{2}{c}{473~nm} \\
300\,K & 5\,K & 295\,K & 10\,K & 300\,K & 5\,K & 300\,K & 5\,K & $E_{2}$(i) $i=1,\dots,13$ (calc.)  & 300\,K & 5\,K & 295\,K & 10\,K & 300\,K & 5\,K & 300\,K & 5\,K \\
\hline
70 & 71 & 71 & 71 & 71 & 73 & 70 & 71 & 71 & 70 & 71 & 71 & 71 & 71 & 72 & 70 & 71 \\
150 & 151 & - & - & - & - & - & - & 148 & 149 & 151 & - & - & - & - & - & - \\
193 & - & 195 & 196 & 194 & 197 & - & - & 191 & 193 & 195 & 195 & 196 & 194 & 196 & - & - \\
215 & 217 & 217 & 219 & 217 & 219 & 216 & 217 & 215 & 216 & 217 & 217 & 219 & 216 & 219 & 216 & 217 \\
276 & 276 & 278 & 278 & 278 & 279 & 277 & 275 & 263 & 277 & 276 & 278 & 278 & 277 & 278 & 276 & 277 \\
337 & 335 & 335 & 337 & 335 & 337 & 337 & 336 & 332 & 333 & 334 & 335 & 337 & 337 & 337 & 335 & 335 \\
348 & 346 & 348 & 347 & 347 & 349 & 349 & 347 & 357 & 348 & 347 & 348 & 347 & 349 & 349 & 347 & 348 \\
- & - & 449 & 450 & 449 & 451 & - & 450 & 458 & 449 & 449 & 449 & 450 & 449 & 452 & 450 & 450 \\
- & - & - & - & 476 & 476 & - & - & 472 & 476 & 475 & - & - & - & - & - & - \\
482 & 483 & 484 & 485 & 484 & 485 & 483 & 483 & 482 & 483 & 483 & 484 & 485 & 484 & 486 & 484 & 482 \\
519 & - & 522 & 522 & 521 & 522 & 520 & 522 & 531 & - & 520 & 522 & 522 & - & 523 & 521 & 521 \\
565 & 566 & 564 & 565 & 566 & 567 & 565 & 566 & 570 & 566 & 566 & 564 & 565 & 565 & 568 & 566 & 566 \\
739 & 739 & 738 & 739 & 738 & 740 & 738 & 739 & 764 & 737 & 737 & 738 & 739 & 737 & 741 & 737 & 738 \\
\hline
- & 208 & 205 & 208 & - & - & - & - & $A_1$(1) at 207 (exp.) & 205 & 207 & 205 & 208 & 206 & 208 & 206 & 207 \\
- & - & 373 & 374 & 372 & 376 & - & 375 & $A_1$(3) at 372 (exp.) & 373 & 373 & 373 & 374 & 373 & 376 & 374 & 374 \\
- & 656 & 654 & 656 & - & 653 & - & 655 & $A_1$(7) at 655 (exp.) & 655 & 657 & 654 & 656 & 654 & 660 & 655 & 657 \\
\hline
- & - & - & 505 & - & 506 & - & - & $B_1$(6) at 497 (calc.) & - & 503 & - & 505 & - & 506 & 502 & 504 \\
- & - & - & - & - & - & - & - & $B_1$(7) at 589 (calc.)  & 578 & 580 & - & - & 578 & 582 & 579 & 580 \\
- & - & - & 820 & - & - & - & - & $B_1$(10) at 840 (calc.) & 820 & 821 & 820 & 820 & 818 & 823 & 819 & 820 \\
\hline
- & 212 & - & - & - & - & - & - & $B_1$(2) at 210 (calc.) & - & - & - & - & - & - & - & - \\
- & 227 & - & 229 & - & 228 & - & - & $E_1$(3) & - & 226 & - & 229 & - & - & - & 227 \\
- & - & - & - & - & - & - & - & unidentified & 285 & - & - & - & - & - & - & 284 \\
- & - & 291 & - & - & - & - & - & $E_1$(4) at 291 (exp.) & 293 & - & 291 & - & - & - & 290 & - \\
- & - & - & 303 & - & - & - & - & electronic & - & 300 & - & 303 & - & 305 & - & 304 \\
- & 361 & - & 361 & - & 362 & - & 361 & electronic & - & 360 & - & 361 & - & 363 & - & - \\
\end{tabular}

\end{ruledtabular}
\end{table*}

Given the large number of expected modes, we start with the nine phonons of $A_1$ symmetry, which should be active in the IR experiment for light polarization $\textbf{E}^\omega\parallel c$ and in the Raman scattering configuration $y(zz)\bar{y}$, where no other vibrational Raman modes should be active. The corresponding reflectivity and Raman spectra are shown in Fig.~\ref{fig:comp_IR_532nm}(a) together with calculated eigenfrequencies (black triangles). At room temperature, we identify eight clearly visible resonances in the IR reflectivity and nine modes in the Raman spectrum, where the lowest-lying mode is very weak in the Raman spectrum and not observable in reflectivity. The number of modes is in good agreement with the expected $A_1$ phonons and they remain visible upon cooling into the magnetically ordered phase. When entering into the magnetic phase only one additional mode emerges at 304~cm$^{-1}$. The eigenfrequencies of all excitations were determined from the peak maxima for the Raman modes and from a fit with Lorentzian lineshapes for the reflectivity spectra (see Fig.~\ref{fig:IR_fit} in Supplement \ref{app:Ramanspectra}) and are listed in Tab.\,\ref{tab:zz} together with the calculated eigenfrequencies for the $A_1$ phonons at low temperatures (the complete list of calculated eigenfrequencies is given in the Tab.~\ref{tab:calculatedfreq} in Supplement \ref{app:calcphonons}). Based on the good agreement of calculated and observed eigenfrequencies of the nine excitations observed at 295~K and 10~K, we identify these modes with the $A_1$ phonons and conclude that we observe no deviations from the approximate selection rules in Eqs.~(\ref{Eq:irreps}) and (\ref{Eq:coreps}). The additional mode at 304~cm$^{-1}$ might be one of the $B_1$ phonons, which are predicted to fall into the same copresentation class as the $A_1$ modes. However, we discard this possibility, as none of the calculated eigenfrequencies of the $B_1$ phonons is in agreement with the observed mode. Therefore, we assign it to be of electronic origin stemming from Co$^{2+}$-multiplet excitations, similar to modes reported for the isostructural material \fmo{} in its magnetically ordered phase \cite{Reschke:2020,Stanislavchuk:2020,Vasin:2024}.

We will follow the same line of reasoning in the assignment of the modes in the other measurement configurations. The richest spectra are observed for infrared reflectivity measured with light polarization $\textbf{E}^\omega\parallel a$ and in the Raman scattering configuration $y(xz)\bar{y}$, which are shown together in Fig.~\ref{fig:comp_IR_532nm}(b) and compared to the calculated eigenfrequencies of the $E_1$ phonons (green triangles). The eigenfrequencies of all observed modes are listed in Tab.~\ref{tab:xz} and we can identify three classes of excitations. 

The first class contains the modes visible at 295~K and at 10~K, which are in good agreement with the expected frequencies of the twelve $E_{1}$(i) phonons $i=1,\dots,12$. Note that modes $E_1(1)$ and $E_1(7)$ are only observed in Raman scattering. The second class contains six modes, which are observed by Raman scattering only, but both at 295~K and at 10~K with eigenfrequencies coinciding with the observed eigenfrequencies of the identified $A_1$ phonons (compare Fig.~\ref{fig:comp_IR_532nm}(a) and Tab.~\ref{tab:zz}). $A_1$ phonons are not allowed in this configuration according to the selection rules and, moreover, their strength is comparable to that of the allowed modes in this configuration. As the appearance is restricted to the Raman spectra only, but they are observed already at 295~K, we attribute their Raman activity and the breaking of the approximate selection rules given in Eqs.~(\ref{Eq:irreps}) and (\ref{Eq:coreps}) to resonant Raman effects. Below we will discuss Raman spectra in $z(xy)\bar{z}$ and $z(xx)\bar{z}$ configuration as a function of the wavelength of the incoming laser beam, which support this interpretation. Moreover, the possibility for resonant Raman effects is further supported by temperature dependent transmission experiments in the mid- and near infrared frequency regime, which are shown in Supplement \ref{app:Ramanspectra}. The spectra reveal the appearance of a fine-structure below $T_\text{N}$ similar to the sister compound \fmo{}, where the excitations were assigned to Fe$^{2+}$ multiplet transitions at energies below the opening of the semiconducting band gap \cite{Stanislavchuk:2020,Vasin:2024}. We believe that this is also the case in \cmo{} and that the mode identified in this work as electronic (see Tables \ref{tab:zz}-\ref{tab:xyresonant}) and the modes in the MIR- and NIR regime can be ascribed to Co$^{2+}$ multiplet transitions. Moreover, the band gap in \cmo{} is determined to be at about 1.4~eV (see Fig.~\ref{fig:absorption}(b) in Supplement \ref{app:Ramanspectra}), indicating that all used Raman laser frequencies can lead to electronic excitations across the band gap and contribute to resonant Raman effects. Note that the observed band gap is in reasonable agreement with the calculated estimate of about 1.55~eV for \cmo{} \cite{Park:2021}.

The third class consists of eight modes, which only emerge upon cooling below the Néel temperature and appear either in both Raman and IR channels or in only one of the channels. Three of these modes can be identified as $E_2$ phonon modes, which are expected to be observable in both the $y(xz)\bar{y}$ and the $z(xy)\bar{z}$ channels in the magnetically ordered phase as a part of the $DE$ copresentation and confirm the expected changes of selection rules due to symmetry breaking by magnetic ordering as derived from the analysis of the relativistic magnetic point group $6^\prime m^\prime m$. Four further modes are assigned to electronic modes, possibly from Co$^{2+}$-multiplet excitations. For the two modes at 87~cm$^{-1}$ and 118~cm$^{-1}$ this interpretation is in agreement with their observed splitting in an applied magnetic field reported in THz transmission measurements \cite{Reschke:2022}. The assignment of the modes at 303~cm$^{-1}$ and 363~cm$^{-1}$ as electronic is based on fact, that there are no calculated matching phonon eigenfrequencies and that they are only emerging in the antiferromagnetic state. Note that the mode at 303~cm$^{-1}$ has also been observed for $\textbf{E}^\omega\parallel c$ and  $y(zz)\bar{y}$ at low temperatures. The mode at 820~cm$^{-1}$ is not considered to be of electronic origin, because it has been observed also at room temperature in the Raman configuration $z(xx)\bar{z}$ (see Tab.~\ref{tab:xyresonant}) for different laser wavelengths. Hence, it is considered to be a Raman active phonon as a result of resonant scattering effects and it is identified with the highest-lying $B_1$ mode by comparison to all calculated eigenfrequencies in Tab.~\ref{tab:calculatedfreq}.

To complete the analyses of Raman selection rules, we show the spectra for the Raman scattering configurations $z(xx)\bar{z}$ and $z(yx)\bar{z}$ in Fig.~\ref{fig:comp_IR_532nm}(c) and (d), respectively, for a laser wavelength of 532~nm (as in panels (a) and (b)) together with the spectra taken at a wavelength of 632.8~nm. Raman spectra taken in these configurations at 514.3~nm and 473~nm are shown in Fig.~\ref{fig:xx_xy_473_514} in Supplement \ref{app:Ramanspectra}. All observed eigenfrequencies for both configurations and all wavelengths are summarized in Tab.~\ref{tab:xyresonant}. Following the selection rules, the spectra in $z(yx)\bar{z}$ configuration should identify the 13$E_2$ phonons at room temperature, while for $z(xx)\bar{z}$ 
both $E_2$ and $A_1$ phonons are allowed at room temperature already. At low temperature all modes of corepresentations $DA_1$ (corresponding to all $A_1$  and $B_1$ modes) and $DE$ (corresponding to all $E_1$ and $E_2$)  are allowed in $z(xx)\bar{z}$, while in $z(yx)\bar{z}$ the observed excitations should be restricted to $DE$.

In comparison with the calculated eigenfrequencies and taking into account all different wavelengths, we can identify all of the 13 expected $E_2$ modes in both configurations and confirm all expected Raman modes for the paramagnetic state. In addition, three of the nine $A_1$ phonons already observed in the $y(xz)\bar{y}$ channel are again present, although their appearance in $z(yx)\bar{z}$  depends on the wavelengths of the incident laser and on temperature. In $z(xx)\bar{z}$ configuration these three modes are observed at both temperatures at all wavelengths, which is reasonable as they are symmetry-allowed in this channel. However, the remaining six allowed $A_1$ phonons are not observed for $z(xx)\bar{z}$, which indicates that the corresponding elements of the $A_1$ polarizability tensor are much smaller than the ones for $y(zz)\bar{y}$. This leads us to conclude that the main Raman activity of these modes stems from resonant effects. 
We also identify a new class of three excitations which appear at both temperatures, as phonons of $B_1$ character in comparison to the calculated eigenfrequencies. The highest lying one at 820~cm$^{-1}$ has also been observed in the  $y(xz)\bar{y}$ channel at low temperature (see Tab.~\ref{tab:xz}). The remaining observed excitations seem to be visible either at RT or at low temperature and, again, the observation depends on the laser wavelength, but a clear pattern seems evasive. Notably, the appearance of mode $E_1(3)$ is restricted to the antiferromagnetic phase, which is in agreement with the expected breaking of the selection rules due to the magnetic symmetry group. 

\begin{figure}[t]
    \centering
    \includegraphics[width=\linewidth]{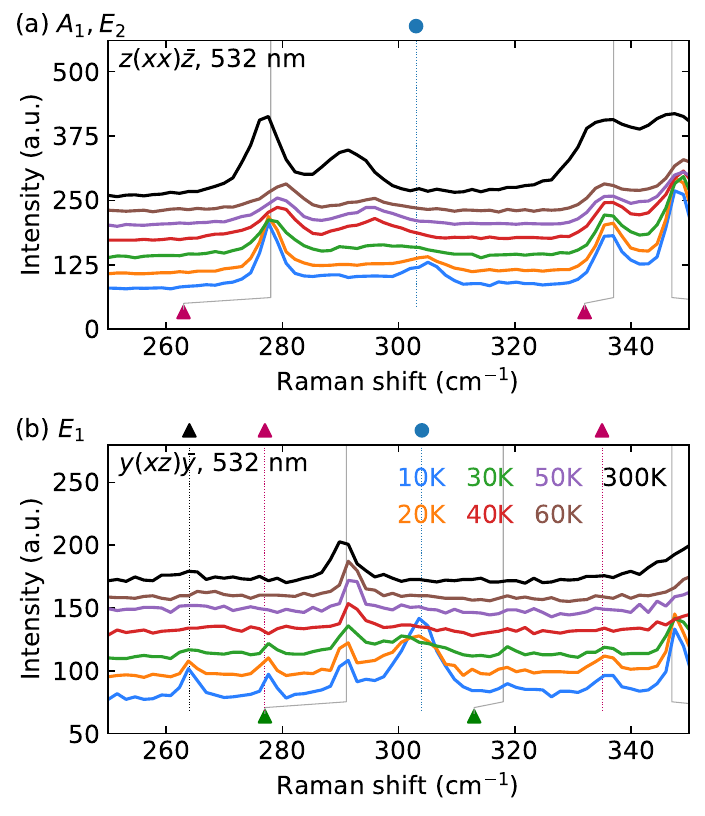}
    \caption{Temperature dependence of the Raman spectra in (a) $z(xx)\bar{z}$ configuration and (b)  $y(xz)\bar{y}$ configuration in the vicinity of the $E_1(4)$ mode at around 290~cm$^{-1}$ and the electronic mode at around 304~cm$^{-1}$. The mode assignment follows the same colorscheme as in Fig.~\ref{fig:comp_IR_532nm}.}
    \label{fig:Tdep}
\end{figure}

In contrast, the observation of mode $E_1(4)$ is restricted to RT only, while it is clearly visible at both temperatures in the $y(xz)\bar{y}$-channel. The temperature dependencies of the two channels in the vicinity of $E_1(4)$ are compared in Fig.~\ref{fig:Tdep} for 532~nm. While in the $y(xz)\bar{y}$-channel
one can clearly see that the mode at 304~cm$^{-1}$ emerges next to $E_1(4)$ below $T_\text{N}$, it seems that in the $z(xx)\bar{z}$ the $E_1(4)$ mode broadens with decreasing temperature and evolves into an enhanced background plateau, while the mode at 304~cm$^{-1}$ emerges.

Moreover, the mode at 360~cm$^{-1}$
is again only visible at low temperature in all but one configuration and wavelength. Two new excitations have been observed, which cannot easily be assigned, one at 212~cm$^{-1}$ only seen at low temperature and only for 632.8~nm, and one at 285~cm$^{-1}$ visible at RT at 632.8~nm and at low temperature at 473~nm. While no calculated eigenfrequency of not yet identified phonons is matching with the latter mode, the former matches well with the calculated $B_1(2)$ eigenfrequency of 210~cm$^{-1}$ and is therefore assigned as such, although it might also correspond to an electronic excitation related to the Co$^{2+}$ multiplet states.

\section{Discussion}
We studied the optical phonons of \cmo{} in its paramagnetic and collinear antiferromagnetic phases by IR- and Raman spectroscopy. We find that (i) we can detect all expected optical phonon modes of the paramagnetic phase in either the IR- or the Raman and, in most cases, both channels, (ii) their eigenfrequencies are in excellent agreement with values obtained by \textit{ab initio} calculations, and most importantly, (iii) the changes of the IR and Raman selection rules induced by the emergence of the antiferromagnetic state follow the predictions derived from irreducible corepresentations of the relativistic magnetic point group. 

In particular, the last two points show that relativistic effects are necessary to capture the essential physics of optical phonons in \cmo. This is in contrast to the derived phonon selection rule for an ideal (spin-orbit free) altermagnetic ground state in \cmo, where the non-relativistic spin group symmetries  predict that phonon selection rules do not change upon the magnetic phase transition. A reanalysis of the IR- and Raman eigenfrequencies published in Refs.~\onlinecite{Reschke:2020, Stanislavchuk:2020} for the sister compound \fmo{} confirms this result. Using the same criteria as for the mode assignment in \cmo{}, we assigned the observed excitations in Tables \ref{tab:zz_fmo}-\ref{tab:xy_fmo} in Supplement \ref{app:calcphonons} and find very similar systematics  with respect to resonant Raman effects at room temperature and to the magnetic symmetry breaking below $T_\text{N}$. In addition, both compounds reportedly show  non-reciprocal directional dichroism \cite{Reschke:2022,Vasin:2024}, which is an optical magnetoelectric effect that is not allowed in the spin group approach. This is consistent with its interpretation as a relativistic correction to the Cotton-Mouton effect \cite{Rikken:2002}. 

Finally, we want to emphasize that our approach to investigate optical phonons for testing symmetry properties of magnetically ordered materials is not constrained to altermagnets but valid for all magnetic compounds associated with the spin group concept such as e.g. $p$-wave magnets \cite{Hellenes:2023,Chakraborty:2024,Song:2025}.

\section{Methods}
\subsection{Reflectivity measurements}
Reflectivity measurements were performed on an as-grown $ab$-plane single crystal and on an $ac$-cut mosaic sample composed of two single crystals. Transmission measurements were performed on a thin $ab$-plane single crystal.  By using a Bruker Fourier-transform IR-spectrometer Vertex80 equipped with a He-flow cryostat, the frequency range from 100 to 13000\,cm$^{-1}$ and a temperature range from 5 to 300\,K could be covered. To determine mode frequencies, fits with Lorentzian oscillators were performed using the RefFIT software \cite{Kuzmenko:2005}.

\subsection{Raman scattering}
Raman scattering spectra were recorded in backscattering geometry using a Jobin Yvon LabRam HR800 micro-spectrometer. A 532\,nm laser with a power of 370\,\(\upmu\)W on the sample was employed as the excitation source with an acquisition time of $120\times3$~s. Focusing was performed using a 50\(\times\) microscope objective. To control the temperature, the samples were placed in a cryostat, which enabled measurements over a wide temperature range. Measurements in $z(yx)\bar{z}$ and  $z(xx)\bar{z}$ configurations were performed on an  \emph{ab}-plane cut single crystal and 
measurement in $y(xz)\bar{y}$ and $y(zz)\bar{y}$ configurations were performed on an \emph{ac}-plane cut single crystal.

 All Raman spectra with 632.8 nm, 514.3 nm, 473 nm wavelengths were recorded using a Trivista 777 spectrometer in single-stage configuration, equipped with a Nitrogen-cooled, ultra-low-noise PyLoN CCD detector. A diffraction grating with 1800 grooves/mm was used, yielding a spectral resolution better than 0.3 cm$^{-1}$. Typical acquisition time was 600 seconds for all the measurements. The measurements were carried out using three different excitation sources: a He–Ne laser with wavelength $\lambda$ = 632.8~nm (1.95~eV), and two  diode-pumped solid-state (DPSS) lasers with $\lambda$ = 514.3~nm (2.41 eV)  and $\lambda$ = 473~nm (2.62~eV). The excitation beam was focused onto the sample using a 50$\times$ long-working-distance objective lens with a numerical aperture (NA) of 0.55, with an approximate spot size of 2 $\upmu$m. To suppress the Rayleigh scattering and achieve a low-energy cutoff, three volume Bragg filters were employed, each optimised for the respective excitation wavelength. The laser power incident on the sample was maintained at approximately 300 $\upmu$W for all measurements. Low-temperature measurements were conducted using a Janis liquid helium (LHe)-based cold-finger cryostat operating under a vacuum of 5 × 10$^{-5}$ mbar. 

\subsection{DFT calculations}
Density-functional-theory (DFT) band-structure calculations were performed in the \texttt{VASP} code~\cite{vasp1,vasp2} using the Perdew-Burke-Ernzerhof version of the exchange-correlation potential~\cite{pbe96}. Phonon frequencies at the $\Gamma$-point were obtained by the finite-displacement method. Electronic correlations in the Co $3d$ shell were taken into account on the mean-field level using the DFT+$U$ procedure with the on-site Coulomb repulsion parameter $U_d=5$\,eV and Hund's coupling $J_d=1$\,eV~\cite{Szaller:2025}. The experimental collinear antiferromagnetic configuration was used, and spin-orbit (SO) coupling was included in order to reproduce the large magnetic anisotropy on the octahedrally coordinated Co$^{2+}$ site.

\section{Data availability}
The IR and Raman spectra are available on Zenodo \href{https://doi.org/10.5281/zenodo.16751740}{10.5281/zenodo.16751740}.

\begin{acknowledgments}
M.K. and K.P. thank Hana Schiff for stimulating discussions. This research was partly funded by the Deutsche Forschungsgemeinschaft (DFG, German Research Foundation)-TRR 360-492547816 and EXC-2123 QuantumFrontiers – 390837967. The support via the project ANCD (cod 011201 Moldova) is also acknowledged. The authors gratefully acknowledge the use of computing resources of the ALCC HPC cluster (Institute of Physics, University of Augsburg). We acknowledge the support of the  LNCMI-CNRS, member of the European Magnetic Field Laboratory (EMFL). This research was supported by the Ministry of Culture and Innovation and the National Research, Development and Innovation Office within the Quantum Information National Laboratory of Hungary (Grant No. 2022-2.1.1-NL-2022-00004) as well as the Hungarian NKFIH Grant No. K 142652 and FK 153003. M.K. was partially funded by the scholarship program DKÖP-25-1-BME-25 of NKFIH and Budapest University of Technology and Economics.
\end{acknowledgments}

\section{Author Contributions}
L.P. and V.T. synthesized and characterized the crystals; F.S. and O.E. measured the reflectivity and transmission spectra; D.U., A.P., C.F. and P.L. measured the Raman spectra; F.S., K.V., O.E. and J.D. analyzed the data; A.A.T. performed the \textit{ab initio} calculations; M.K., K.P., S.B. and J.D. performed the group theoretical analysis; F.S., M.K., K.P., S.B. and J.D. wrote the paper; I.K., S.B. and J.D. planned and coordinated the project; All authors contributed to the discussion and interpretation of the experimental and theoretical results and to the completion of the paper.

\section{Competing Interests}
The authors declare no competing interests.

\bibliography{Bibliography}
\clearpage

\beginsupplement
\section{Detailed Group-theoretical considerations for \cmo}
\subsection{Classifications and tables}\label{app:grouptheory}
Here, we review the magnetic point group and spin group classifications of \cmo{}.

\subsubsection{The magnetic point group}
\begin{table}[b]
\caption{Character table of the point group $6mm$.}
    \label{tab:6mm}
\[
\begin{array}{c|rrrrrr}
6mm & E & 2C_6 & 2C_3 & C_2 & 3m_d & 3m_v \\ 
\hline
A_1  & 1 & 1 & 1 & 1 & 1 & 1 \\ 
A_2  & 1 & 1 & 1 & 1 & -1 & -1 \\ 
B_1  & 1 & -1 & 1 & -1 & -1 & 1 \\ 
B_2  & 1 & -1 & 1 & -1 & 1 & -1 \\ 
E_1  & 2 & 1 & -1 & -2 & 0 & 0 \\ 
E_2  & 2 & -1 & -1 & 2 & 0 & 0 \\ \hline
\end{array}
\]
\end{table}

When magnetic order emerges continuously, i.e.~across a second-order phase transition, the magnetic point group $\mathbf{M}$ can be formally constructed from the crystallographic point group $\mathbf{G}$ by identifying a halving subgroup $\mathbf{H}$ of $\mathbf{G}$ and combining the remaining half of the elements of $\mathbf{G}$ with the anti-unitary time-reversal operation $\tau$ (see Ref.~\onlinecite{Bradley:2009}):
\begin{equation}
    \mathbf{M}=\mathbf{H}+\tau(\mathbf{G}-\mathbf{H})
\end{equation}
As the choice of the halving subgroup $\mathbf{H}$ is not unique, a single crystallographic point group $\mathbf{G}$ can give rise to multiple magnetic point groups $\mathbf{M}$.
For the case of \cmo{}, the point group is $\mathbf{G}=6mm$ in the paramagnetic phase, with the character table given in Tab.~\ref{tab:6mm}. The halving subgroup (character table in Tab.~\ref{tab:3m})
\begin{table}[t]
\caption{Character table of the halving unitary subgroup $3m$.}
    \label{tab:3m}
\[
\begin{array}{c|rrr}
3m & E & 2C_3 & 3m_v \\ 
\hline
A_1  & 1 & 1 & 1 \\ 
A_2  & 1 & 1 & -1 \\ 
E  & 2 & -1 & 0 \\ 
\hline
\end{array}
\]
\end{table}
\begin{equation}
  \mathbf{H}=\{E,C_3,m_v\}=3m \quad (\equiv C_{3v})
\end{equation}
leads directly to the correct magnetic point group:
\begin{equation}
    \mathbf{M}=\{E,C_3,m_\mathrm{v},
    C^{\prime}_6,C^\prime_{2},m_\mathrm{d}^\prime\},
\end{equation}
commonly denoted as $6^\prime m^\prime m$  (magnetic point group no. 47 \cite{Bradley:2009}). Here, the symmetry operations $C^{\prime}_6$ and $m_v$ remain non-symmorphic (they are each combined with a $c/2$ translation along $z$), resulting in the magnetic space group $P6^\prime_3 m^\prime c$. The primes indicate symmetry operations combined with time reversal, e.g. $C^{\prime}_6 = \tau C_6$. Practically, the magnetic point group can be determined by checking which of the symmetry operations of the crystallographic point group  (in our case $\mathbf{G}$) need to be combined with the time-reversal operation to leave the magnetically ordered state invariant. In a Lorentz-invariant relativistic framework, the simultaneous transformation of spatial coordinates and spin degrees of freedom naturally arises from the spin-orbit coupling.

Note that, in principle, the point groups describing the paramagnetic state are formally known as grey groups given by $\mathbf{G}+\tau \mathbf{G}$. In the absence of magnetic ordering, the time reversal becomes a symmetry of the grey groups and commutes with all elements of $\mathbf{G}$ \cite{Bradley:2009}. Hence, it is common practice to use only the crystallographic point group $\mathbf{G}$ when discussing the symmetry properties of the paramagnetic state.

\subsubsection{The spin group}
 \begin{table*}[htb]
      \caption{Irreducible corepresentations of the spin group $^{\bar{1}}6^{1}m^{\bar{1}}m$. In this table, we provide the matrices for the generators, from which the remaining elements can be calculated. For reasons of readability, we apply the abbreviations $r=\mathrm{e}^{\mathrm{i}\phi}$ and $s=\mathrm{e}^{2\mathrm{i}\pi / 3}$, and the overbar indicates complex conjugation.}
      \label{tab:char-spingr6mm}
\begin{ruledtabular}
      \begin{tabular}{ccccccc}
         & $[R_\phi || E]$ 
         & $[2_\perp R_\phi || C_{6z} ]$ 
         & $[2_\perp R_\phi||m_{xz}]$ 
         & $[\tau\,2_\perp R_\phi||E]$ 
         & $[\tau\,R_\phi||C_{6z}]$ 
         & $[\tau\,R_\phi || m_{xz}]$ \\
        \hline
        
        $\Gamma_1$ & 1 & 1 & 1 & 1 & 1 & 1 \\
        $\Gamma_2$ & 1 & $-1$ & $-1$ & 1 & $-1$ & $-1$ \\
        $\Gamma_3$ & 1 & 1 & $-1$ & 1 & 1 & $-1$ \\
        $\Gamma_4$ & 1 & $-1$ & 1 & 1 & $-1$ & 1 \\

        $\Gamma_5$ &
        $\begin{pmatrix}1&0\\0&1\end{pmatrix}$ &
        $\begin{pmatrix}{\bar s}&0\\0&s\end{pmatrix}$ &
        $\begin{pmatrix}0&1\\1&0\end{pmatrix}$ &
        $\begin{pmatrix}0&1\\1&0\end{pmatrix}$ &
        $\begin{pmatrix}0&s\\{\bar s}&0\end{pmatrix}$ &
        $\begin{pmatrix}1&0\\0&1\end{pmatrix}$ \\

        $\Gamma_6$ &
        $\begin{pmatrix}1&0\\0&1\end{pmatrix}$ &
        $\begin{pmatrix}-{\bar s}&0\\0&-s\end{pmatrix}$ &
        $\begin{pmatrix}0&-1\\-1&0\end{pmatrix}$ &
        $\begin{pmatrix}0&1\\1&0\end{pmatrix}$ &
        $\begin{pmatrix}0&-s\\-{\bar s}&0\end{pmatrix}$ &
        $\begin{pmatrix}-1&0\\0&-1\end{pmatrix}$ \\

        $\Gamma_7$ &
        $\begin{pmatrix}r^{\nu}&0\\0&{\bar r}^{\nu}\end{pmatrix}$ &
        $\begin{pmatrix}0&{\bar r}^{\nu}\\r^{\nu}&0\end{pmatrix}$ &
        $\begin{pmatrix}0&{\bar r}^{\nu}\\r^{\nu}&0\end{pmatrix}$ &
        $\begin{pmatrix}r^{\nu}&0\\0&{\bar r}^{\nu}\end{pmatrix}$ &
        $\begin{pmatrix}0&{\bar r}^{\nu}\\r^{\nu}&0\end{pmatrix}$ &
        $\begin{pmatrix}0&{\bar r}^{\nu}\\r^{\nu}&0\end{pmatrix}$ \\

        $\Gamma_8$ &
        $\begin{pmatrix}r^{\nu}&0\\0&{\bar r}^{\nu}\end{pmatrix}$ &
        $\begin{pmatrix}0&{-\bar r}^{\nu}\\-r^{\nu}&0\end{pmatrix}$ &
        $\begin{pmatrix}0&{-\bar r}^{\nu}\\-r^{\nu}&0\end{pmatrix}$ &
        $\begin{pmatrix}r^{\nu}&0\\0&{\bar r}^{\nu}\end{pmatrix}$ &
        $\begin{pmatrix}0&{-\bar r}^{\nu}\\-r^{\nu}&0\end{pmatrix}$ &
        $\begin{pmatrix}0&{-\bar r}^{\nu}\\-r^{\nu}&0\end{pmatrix}$ \\

        $\Gamma_9$ &
        $\begin{pmatrix}
         r^{\nu}&0&0&0\\
         0&r^{\nu}&0&0\\
         0&0&{\bar r}^{\nu}&0\\
         0&0&0&{\bar r}^{\nu}
         \end{pmatrix}$ &
        $\begin{pmatrix}
         0&0&s{\bar r}^{\nu}&0\\
         0&0&0&{\bar s}{\bar r}^{\nu}\\
         r^{\nu}&0&0&0\\
         0&r^{\nu}&0&0
         \end{pmatrix}$ &
        $\begin{pmatrix}
         0&0&0&s{\bar r^{\nu}}\\
         0&0&{\bar s}{\bar r^{\nu}}&0\\
         0&sr^{\nu}&0&0\\
         {\bar s}r^{\nu}&0&0&0
         \end{pmatrix}$ &
        $\begin{pmatrix}
         0&r^{\nu}&0&0\\
         r^{\nu}&0&0&0\\
         0&0&0&{\bar r}^{\nu}\\
         0&0&{\bar r}^{\nu}&0
         \end{pmatrix}$ &
        $\begin{pmatrix}
         0&0&0&s{\bar r}^{\nu}\\
         0&0&{\bar s}{\bar r}^{\nu}&0\\
         0&r^{\nu}&0&0\\
         r^{\nu}&0&0&0
         \end{pmatrix}$ &
        $\begin{pmatrix}
         0&0&s{\bar r^{\nu}}&0\\
         0&0&0&{\bar s}{\bar r^{\nu}}\\
         sr^{\nu}&0&0&0\\
         0&{\bar s}r^{\nu}&0&0
         \end{pmatrix}$ 
      \end{tabular}
      \end{ruledtabular}
    \end{table*}
Let us now turn to the case of zero spin-orbit coupling and the description of \cmo{} in terms of spin groups. The spin groups were introduced and elaborated in the 1960s and 1970s \cite{Litvin:1974,Litvin1977}. Here we adopt Litvin’s notation and classification from Ref.~\onlinecite{Litvin1977}. A recent overview on spin groups and their (co-)representations was given by Schiff \textit{et al.}~ \cite{Schiff:2025} and others \cite{Xiao:2024,Chen:2024,Watanabe:2024}. This non-relativistic framework has gained renewed interest because it naturally describes altermagnets — compensated antiferromagnets whose band structures exhibit splittings unrelated to spin-orbit coupling, i.e. lifting of the two-fold Kramers degeneracy along general directions in reciprocal space. More precisely, the time reversal $\tau$ enforces $\varepsilon_\uparrow(\mathbf{k})=\varepsilon_\downarrow(-\mathbf{k}$), the spatial inversion $I$ enforces $\varepsilon_\uparrow(\mathbf{k})=\varepsilon_\uparrow(-\mathbf{k})$. Taken together, they impose $\varepsilon_\uparrow(\mathbf{k})=\varepsilon_\downarrow(\mathbf{k})$ throughout the Brillouin zone, leading to the Kramers degeneracy in a conventional antiferromagnet.
In an altermagnet, however, $I$ and $\tau$ do not simultaneously remain symmetries, so this protection is lifted.

By these criteria, \cmo{} qualifies as an altermagnet: in its antiferromagnetic phase neither spatial inversion nor time reversal is a symmetry, the only operations linking the two collinear sublattices are the nonsymmorphic screw axes ($C^{\prime}_6$) and glide planes ($m_\mathrm{v}$), which never map a generic crystal momentum $\mathbf{k}$ to $-\mathbf{k}$. 
Consequently, no symmetry enforces Kramers degeneracy at generic $\mathbf{k}$, allowing spin‐splittings to appear in the band structure.
In a Mott insulator, these electronic splittings lie at high energies, but the same symmetry arguments apply equally to other quasiparticles -- most notably magnons \cite{McClarty:2024,Schiff:2024}. 
  
In the following, we discuss the spin point group of \cmo{} in its ordered (collinear) phase, before comparing the resulting consequences for observable vectorial or tensorial quantities, such as polarization or the Raman polarizability tensor, with the results of the magnetic point group.

In the spin‐group formalism, the pure spin symmetry of a collinear magnet such as \cmo{} is described by the “spin‐only” group
\begin{equation}
    \mathbf{b}^{\infty} = \text{SO(2)}_{\parallel} \rtimes \{ E^s, \tau C^s_{2\perp} \},
\end{equation}
where $\text{SO(2)}_{\parallel}$ includes all continuous spin rotations about the ordered‐moment axis, while the two‐element subgroup $\{ E^s, \tau C^s_{2\perp} \}$ contains the identity $E^s$ and the antiunitary operation formed by a $\pi$-rotation in spin space about any axis perpendicular to the spin direction, followed by time reversal $\tau$. 
These operations act solely on spin space—leaving the lattice positions unchanged—and reflect the orientations of the spins. In this terminology, the spins form an axial‐vector field embedded in the crystal lattice.

 The spin point group $\mathbf{G}_s$ combines the real‐space symmetries of the crystal with the pure spin operations in $\mathbf{b}^{\infty}$. By factoring out the latter, we get the finite 
 %spin parental spin group $\mathbf{B}=%
quotient group $\mathbf{G}_s/\mathbf{b}^{\infty}$ that is isomorphic to the magnetic point group
$\mathbf{M}$ (which for \cmo{} is $6^\prime m^\prime m$).
 Hence, the spin point group is given by 
\begin{equation}
    \mathbf{G}_s\cong\mathbf{b}^{\infty} \times \mathbf{M}.
\end{equation}
Equivalently, writing $\mathbf{H}$ for the halving subgroup of $\mathbf{G}$
\begin{multline}
  \mathbf{G}_s \cong \{ \left[ b \beta (h) || h  \right] | b \in 
\mathbf{b}^{\infty}, h \in \mathbf{H}\} \\ \cup
     \{ \left[ b \beta (h') || h' \right] | b \in \mathbf{b}^{\infty}, h' 
\in (\mathbf{G}- \mathbf{H})\} ,
     \label{eq:Gs}
\end{multline}
where $\left[ s || g\right]$  denotes the operation that applies $s$ in spin space and $g$ in real space, i.e., axial vector components are transformed with operation $b$ and real space coordinates with $g$. Operations $\beta (h)$ and $\beta (h')$ are the associated spin space transformation of $h$ and $h'$ in the magnetic space group $\mathbf{M}$. Importantly, $\beta{h}$ refers to a unitary operator, while $\beta{h'}$ to an antiunitary one. For example, $\beta (m_{110})$ (referring to a glide plane) is a 2-fold unitary rotation along the (110) direction, while $\beta (m_{100})$ (a diagonal reflection) is the time reversal combined with a 2-fold rotation along the (100) direction. In the specific case of \cmo{}:
\begin{align}
    \mathbf{G}_s & \cong 
    \left( \text{SO(2)}_{\parallel} 
    \rtimes \{ E^s, \tau C^s_{2\perp} \} \right)
    \times 6^\prime m^\prime m 
    \nonumber\\
    & = ^{\bar{1}}\!6^{\bar{1}}m^1m
\end{align}
in Litvin's notation \cite{Litvin1977}.
% need to ask Mark
Here, $\bar{1}$ denotes a collinear magnetic arrangement. In this case, the symmetry elements $b$ in Eq.~(\ref{eq:Gs}) are combined with the identity when acting within the sublattice (the first term) and with the pure time-reversal when exchanging magnetic sublattices (the second term). Table \ref{tab:char-spingr6mm} lists the irreducible corepresentations of the spin group $^{\bar{1}}6^{1}m^{\bar{1}}m$ for \cmo{} taken from Schiff \textit{et al}.~\cite{Schiff:2025}.

Let us note that this notation is not universally adopted. For example, in Ref.~\onlinecite{Smejkal:2024}, the spin group of \fmo{} (which has the same symmetry group as \cmo{}) is written as $^26^2m^1m$.
In Litvin's notation, however, superscripts different from $\bar{1}$ indicate non-collinear spins: the $^26^2m^1m$ would represent a coplanar magnetic structure (and $^m6^mm^1m$ for a non-coplanar one). 

\subsection{Framework of (co)representations for IR- and Raman-active modes}\label{app:coreps}

Let us begin by determining the selection rules for IR-active phonons. For this, we need to consider transition matrix elements induced by the interaction between phonons and the electric field of incident light. 
The relevant matrix element is expressed as $\langle f|\mathbf{E}^{\omega}\cdot\mathbf{p}|i\rangle$, where $\mathbf{E}^{\omega}$ denotes the electric field vector of the incident light, $\mathbf{p}$ is the electric-dipole moment operator associated with harmonic lattice vibrations, and $|i\rangle$ and $|f\rangle$ represent the initial and final phonon states, respectively.
Within the harmonic approximation, the phonon states form representations of the underlying symmetry group. 
Consequently, initial phonon states can be labeled by representation $\Gamma_i$ and final by  $\Gamma_f$. 
The electric-dipole operator $\mathbf{p}$, arising directly from lattice vibrations, transforms as a polar vector described by the representation $\Gamma_P$, and we expect the electric field $\mathbf{E}^{\omega}$ to also transform as a polar vector (i.e. $\Gamma_{\mathbf{E}^{\omega}} \equiv \Gamma_P$).
A finite, non-zero expectation value of the matrix element requires that it transforms as the trivial (identity) representation under all symmetry operations of the group. Equivalently, the direct product representation 
$\Gamma_f^*\times\Gamma_{\mathbf{E}^{\omega}}\times\Gamma_{\mathbf{p}}\times\Gamma_i \equiv \Gamma_f^*\times\Gamma_{P}\times\Gamma_{P}\times\Gamma_i$ must include the trivial irreducible representation $\Gamma_1$, ensuring the invariance of the matrix element under unitary symmetry transformations \cite{Tinkham:1964}. 
This approach can be extended to magnetic groups by employing corepresentations to account for antiunitary operations, such as time-reversal symmetry \cite{Bradley:2009}. Importantly, for IR spectroscopy involving electric-dipole transitions, the selection rules derived from the non-magnetic crystallographic point group $\mathbf{G}$ and the paramagnetic (grey) magnetic point group $\mathbf{G}+\tau \mathbf{G}$ coincide, since the perturbation $\mathbf{E}^{\omega}\cdot \mathbf{p}$ remains invariant under time reversal.

Let us now turn to the Raman scattering cross section. In principle, the transformation properties of the Raman polarizability tensor are not straightforward. When the phonon frequency is negligible compared to those of the incident and scattered light -- as realized in our experiment -- the tensor simplifies to a second‐rank symmetric form  (see Refs.~\onlinecite{Birman:1974,Hayes:2004}). This first-order non-resonant Raman cross section for non-polar phonons is given by
$\sigma_{R}\propto \left|\mathbf{e}^*_\text{Sc}\cdot\hat{R}\cdot\mathbf{e}^{\phantom{*}}_\text{In}\right|^2$, with the reduced Raman polarizability tensors $\hat{R}$ and the polarization vectors $\mathbf{e}_\text{In}$ and $\mathbf{e}_\text{Sc}$ of the incoming and scattered light, respectively \cite{Hayes:2004}. The polarization vectors $\mathbf{e}=\mathbf{E}^{\omega}/|\mathbf{E}^{\omega}|$ naturally transform like polar vectors with representation $\Gamma_P$. For the entire cross section to be invariant under the symmetry operations of the crystal, the direct product $\Gamma_P^*\times\Gamma_R\times\Gamma_P$ must contain the identical representation $\Gamma_1$. Again, this ensures that the scattering cross section is invariant under any unitary transformation. Raman tensors for various point groups are tabulated in Refs.~\onlinecite{Hayes:2004, Cardona:1982}, and the corresponding symmetry-adapted functions enable identification of the allowed scattering geometries for observing specific phonon modes.

In Tab.~\ref{tab:Coreps_cmo} we schematically show the relation of irreducible (co)representation between the grey point group $\mathbf{g}$, the crystallographic point group $\mathbf{G}$, its halving subgroup $\mathbf{H}$, and the magnetic point group $\mathbf{M}$ of \cmo{} following the scheme by Anastassakis and Burstein \cite{Anastassakis:1972}.  The characters of $A_1,A_2$ are clearly the same for the conjugacy classes in both groups, $B_1$ has the same characters as $A_1$ and $B_2$ the same as $A_2$.
The irreducible representation $E_1,E_2$ have the same characters as $E$. As a result the irreps $B_1,B_2$ are reduced to $A_1,A_2$, respectively, and $E_1,E_2$ are reduced to $E$. According to the rules in Ref.~\onlinecite{Bradley:2009}, the corresponding irreducible corepresentation of the magnetic point group $6^\prime m m^\prime$ are of type (a), i.e. we can simply imply $\Gamma \rightarrow D\Gamma$.

\begin{table}[t]
 \caption{Reduction of the irreducible representations of the crystallographic point group $\mathbf{G}=6mm$ with respect to the unitary halving subgroup $\mathbf{H}=3m$ to determine the corepresentations of the magnetic point group $\mathbf{M}=6^\prime  m m^\prime$.}
    \label{tab:Coreps_cmo}
    \centering
    \begin{tabular}{c|c|c|c}
      $\mathbf{g}=\mathbf{G} \oplus\{E+\tau\}$   & $\mathbf{G}$ & $\mathbf{H}$ &  $\mathbf{M}=\mathbf{H}+\tau(\mathbf{G}-\mathbf{H})$ \\
         & $6mm$ & $3m$ &  $6^\prime m m^\prime$ \\\hline
    DA$_1$  &  A$_1$ & A$_1$ & DA$_1$\\
    DA$_2$  &  A$_2$ & A$_2$ & DA$_2$\\
    DB$_1$  &  B$_1$ & A$_1$ & DA$_1$\\
    DB$_2$  &  B$_2$ & A$_2$ & DA$_2$\\
    DE$_1$  &  E$_1$ & E &  DE\\
    DE$_2$  &  E$_2$ & E &  DE\\
    \end{tabular}
   
\end{table}

The reduced Raman tensors for calculating the scattering cross section $\sigma_{R}\propto \left|\mathbf{e}^*_\text{Sc}\cdot\hat{R}\cdot\mathbf{e}^{\phantom{*}}_\text{In}\right|^2$ are given in Tab.~\ref{tab:Ramantensors} for the irreducible representations of $6mm$ and the corepresentations of $6^\prime m^\prime m$ \cite{Hayes:2004,Cracknell:1969}. 
\begin{table}[]
        \centering
        \caption{Raman tensors of the irreducible (co-)representations for \cmo.}
        \begin{tabular}{l  l}
            \textbf{\( A_1: \)}  &  \textbf{\( DA_1: \)} \\
             \quad$
             \begin{bmatrix}
                a & 0 & 0 \\
                0 & a & 0 \\
                0 & 0 & b
            \end{bmatrix}
            $ & \quad$
            \begin{bmatrix}
                A & 0 & 0 \\
                0 & A & 0 \\
                0 & 0 & B
            \end{bmatrix}
            $\\
            & \\
            \textbf{\( A_2: \)}  &  \textbf{\( DA_2: \)} \\
             \quad$
             \begin{bmatrix}
                0 & c & 0 \\
                -c & 0 & 0 \\
                0 & 0 & 0
            \end{bmatrix}
            $ & \quad$
            \begin{bmatrix}
                0 & C & 0 \\
                -C & 0 & 0 \\
                0 & 0 & 0
            \end{bmatrix}
            $\\
            & \\
            \textbf{\( E: \)}  &  \textbf{\( DE: \)} \\
             $
             \begin{bmatrix}
                0 & d & 0 \\
                d & 0 & e \\
                0 & f & 0
            \end{bmatrix}, \quad
            \begin{bmatrix}
                d & 0 & -e \\
                0 & -d & 0 \\
                -f & 0 & 0
            \end{bmatrix}
            $ & $
            \begin{bmatrix}
                0 & F & 0 \\
                F & 0 & iD \\
                0 & iE & 0
            \end{bmatrix}, \quad
            \begin{bmatrix}
                -F & 0 & iD \\
                0 & F & 0 \\
                iE & 0 & 0
            \end{bmatrix}
            $\\
        \end{tabular}
        \label{tab:Ramantensors}
\end{table}

\subsection{Transformation of physical quantities under spin group symmetries}\label{app:light_transform}

First, let us take a generic spin group symmetry $\left[ s || g \right] $. Following the conventions of Ref.~\onlinecite{Schiff:2025}, this symmetry element acts on the real space vectors as 
\begin{equation}
    \left[ s || g \right] \mathbf{r} = g\mathbf{r}
\end{equation} 
and on the magnetization field as
\begin{equation}
    \left[ s || g \right] \mathbf{m}(\mathbf{r}) = s\mathbf{m}(g^{-1}\mathbf{r}).
\end{equation} 
This example illustrates the importance of distinguishing between two types of vectors: (i) Purely spatial vectors, which transform trivially under the spin operation $s$ but nontrivially under the spatial part $g$; (ii) Particular embedded vector fields (like $\mathbf{m}$), whose components are transformed by the spin space operations in addition to transforming the spatial $\mathbf{r}$ by the action of $g$.

Since an electric-dipole moment $\mathbf{p}$ localized at an atomic position $\mathbf{r}$ originates from the real-space displacements of the electric charges, it behaves as a true polar vector under spatial rotations. Hence for any spin-space element 
$\left[ s || g \right]$,
\begin{equation}
    \left[ s || g \right] \mathbf{p}(\mathbf{r}) = g\mathbf{p}(g^{-1}\mathbf{r});,
\end{equation}
with no action by the spin‐rotation $s$.
Since the Raman polarizability tensor $\hat{R}$ is constructed as a symmetric dyadic product of these local dipoles, $\hat{R} \sim \mathbf{p} \circ \mathbf{p}$, it remains invariant under spin rotations as well.  In other words, both $\mathbf{p}$ and $\hat{R}$ transform non-trivially only under the spatial part $g$ of the spin group and are unaffected by the spin operation $s$.

In IR spectroscopy, we probe the matrix elements of the light–matter interaction 
\begin{equation}\mathbf{E}^\omega \cdot \mathbf{p} \,,
\end{equation} 
which is allowed by symmetry only if the electric field
$\mathbf{E}^\omega$ and the electric-dipole moment $\mathbf{p}$ transform identically. Accordingly, under a spin–group operation $\left[ s || g \right]$ we take
\begin{equation}
 \left[ s || g \right] \mathbf{E}^\omega(\mathbf{r}) = g\mathbf{E}^\omega(g^{-1}\mathbf{r}) \;,
\end{equation}
with the spin rotation $s$ acting trivially on $\mathbf{E}^\omega$.
In the non-relativistic limit this reproduces the usual selection rules of the paramagnetic grey point group.

However, the arguments above contradict the relativistic nature of the light. Namely, the Zeeman coupling 
$\mathbf{m} \cdot \mathbf{H}^\omega$
between local magnetization $\mathbf{m}$ and the magnetic field $\mathbf{H}^\omega$ of the light requires that both  $\mathbf{m}$ and $\mathbf{H^\omega}$ transform identically, meaning 
\begin{equation}
    \left[ s || g \right] \mathbf{H}^\omega(\mathbf{r}) = s\mathbf{H}^\omega(g^{-1}\mathbf{r})\;.
\end{equation}
Since the vector product of $\mathbf{H}^\omega$ and $\mathbf{E}^\omega$ describes a real-space propagation vector $\mathbf{k} = \mathbf{E}^\omega \times \mathbf{H}^\omega$, it must satisfy
\begin{equation}
    \left[ s || g \right] \mathbf{k} = g \mathbf{k} = g (\mathbf{E}^\omega \times \mathbf{H}^\omega)\;.
\end{equation}
A naive combination of trivial spin action on 
$\mathbf{E}^\omega$ with non-trivial action on 
$\mathbf{H}^\omega$ would give
\begin{equation}
    \left[ s || g \right] \mathbf{k} = (g \mathbf{E}^\omega) \times (s \mathbf{H}^\omega) \neq g (\mathbf{E}^\omega \times \mathbf{H}^\omega) \;,
\end{equation}
revealing an apparent inconsistency.
Thus, while we may still exploit the spin‐group to constrain the dominant electric‐dipole terms in IR and Raman scattering, we must remain mindful of the underlying relativistic coupling between $\mathbf{E}^\omega$, $\mathbf{H}^\omega$, and $\mathbf{k}$ when deriving rigorous selection rules.

\clearpage

\section{Calculated phonon eigenfrequencies for \protect\cmo{} and \protect\fmo{} and comparison for \protect\fmo{}} \label{app:calcphonons}
In Tab.~\ref{tab:calculatedfreq} we list all calculated eigenfrequencies obtained from density-functional-theory (DFT) band-structure calculations for \cmo{} and the sister compound \fmo{}, performed using the experimental collinear antiferromagnetic configuration and including spin-orbit (SO) coupling. Note that the values for \fmo{} were published previously in Ref.~\onlinecite{Reschke:2020}. These values are used for the identification of the experimentally observed modes in both compounds.

\begin{table}[b]
\caption{ \label{tab:calculatedfreq} Calculated phonon eigenfrequencies (in cm$^{-1}$) of the low-temperature antiferromagnetic phase ($T=1.7$~K) in \cmo{} (CMO) and \fmo{} (FMO). Data for \fmo{} is taken from  \cite{Reschke:2020}.}
\centering
\begin{ruledtabular}
\begin{tabular}{cc|cc|cc}
CMO & FMO  &CMO & FMO  & CMO & FMO  \\\hline
\multicolumn{2}{c}{$A_1/DA_1$} & \multicolumn{2}{c}{$A_2/DA_2$} & \multicolumn{2}{c}{$E_1/DE$} \\\hline
204 & 201    & 139 & 141   & 166 & 162 \\
271 & 262    & 403 & 403   & 186 & 191\\
356 & 365    & 446 & 434   & 224 & 223\\
453 & 444    &     &       & 277 & 286 \\
456 & 454    &     &       & 313 & 312 \\
564 & 558    &     &       & 355 & 351 \\
652 & 651    &     &       & 463 & 455 \\
723 & 734    &     &       & 467 & 473\\
811 & 787    &     &       & 480 & 481 \\
    &        &     &       & 528 & 522 \\
    &        &     &       & 575 & 577 \\
    &        &     &       & 760 & 750 \\
        \hline
\multicolumn{2}{c}{$B_1/DA_1$} & \multicolumn{2}{c}{$B_2/DA_2$} & \multicolumn{2}{c}{$E_2/DE$} \\\hline
160 & 155 & 141 & 143 & 71 & 77 \\
210 & 208 & 412 & 413 & 148 & 147 \\
248 & 253 & 440 & 427 & 191 & 191 \\
349 & 363 &      &      & 215 & 215 \\
450 & 449 &      &      & 263 & 277 \\
497 & 481 &      &      & 332 & 334 \\
589 & 586 &      &      & 357 & 359 \\
646 & 647 &      &      & 458 & 453 \\
732 & 738 &      &      & 472 & 472 \\
840 & 815 &      &      & 482 & 482 \\
    &        &     &       & 531 & 525 \\
    &        &     &       & 570 & 572 \\
    &        &     &       & 764 & 751 \\
\end{tabular}
\end{ruledtabular}
\end{table}

It becomes clear from the comparison, that the eigenfrequencies are very close for both compounds as expected for two isostructural materials with the same magnetic structure. Since the agreement of calculated and experimental phonon eigenfrequencies for \cmo{} is very good, we reanalyzed the corresponding published data \cite{Reschke:2020,Stanislavchuk:2020,Wu:2023} for \fmo{} and list the corresponding modes in the following tables in comparison with calculated eigenfrequencies \cite{Reschke:2020}, too. 

For the $A_1$ modes in Tab.~\ref{tab:zz_fmo}, very similar behavior as in \cmo{} is observed, including good agreement between theory and experiment and that the lowest lying $A_1(1)$ mode is not visible in reflectivity while its Raman cross section seems to be even smaller than in \cmo{} as it was also not observed in the Raman channel. In addition to three unidentified modes present above and below the antiferromagnetic transition, one additional mode of the magnetically ordered phase at 230~cm$^{-1}$ is identified as an electronic excitation in agreement with calculated Fe$^{2+}$ multiplet states reported by Vasin \textit{et al}.\,\cite{Vasin:2024}.

\begin{table}[t]
\caption{\label{tab:zz_fmo}
Comparison of experimental IR- and Raman-active  excitation frequencies (in cm$^{-1}$) in \fmo{} measured for light polarization \Epara{} \cite{Reschke:2020} and in the Raman configuration $y(zz)\bar{y}$ \cite{Stanislavchuk:2020}, respectively. The experimental eigenfrequencies measured above and below the magnetic phase transition are compared with theoretical \textit{ab initio} values for the nine expected $A_1$ modes and calculated Fe$^{2+}$ multiplet states reported by Vasin \textit{et al}.\,\cite{Vasin:2024}.}
\begin{ruledtabular}
\begin{tabular}{cc|cc|c}
\multicolumn{2}{c|}{IR\cite{Reschke:2020}}          & \multicolumn{2}{c|}{Raman \cite{Stanislavchuk:2020}}   &  Mode \\
\multicolumn{2}{c|}{\Epara{}}   & \multicolumn{2}{c|}{$y(zz)\bar{y}$}  & assignment       \\
70\,K & 5\,K                  & 85\,K & 5\,K               & $A_{1}$(i) $i=1,\dots,9$  (calc.)  \\
\hline
 -   &  -   & - & - & 201 \\
269  & 269  & 260 & 263 & 262 \\
371  & 371  & 368 & 369 & 365 \\
447  & 446  & 445 & 446 & 444 \\
458  & 457  & 453 & 453 & 454 \\
558  & 556  & 553 & 553 & 558 \\
643  & 643  & 643 & 644 & 651 \\
727  & 727  & 724 & 734 & 734 \\
782  & 782  & 769 & 771 & 787 \\
\hline
 -   & 230  &  -  & 232 &  electronic at 233 (calc.)  \\
  -   &  -   & 668 & 668 & not identified \\
    831   &  830   & - & - & not identified \\
  857   &  856   & - & 852 & not identified \\
\end{tabular}
\end{ruledtabular}
\end{table}

\begin{table}[h]
\caption{\label{tab:xz_fmo}
Comparison of experimental IR- and Raman-active  excitation frequencies (in cm$^{-1}$) in \fmo{} measured  above and below the magnetic phase transition for light polarization $\mathbf{E}^\omega\parallel a$ \cite{Reschke:2020} and in the Raman configuration $y(xz)\bar{y}$ \cite{Stanislavchuk:2020}, respectively. 
Mode assignment is made by comparison with calculated phonon eigenfrequencies for $E_1$ and $B_1$ modes and by comparing to experimental values from other Raman measurement configurations and calculated Fe$^{2+}$ multiplet states reported by Vasin \textit{et al}.\,\cite{Vasin:2024}. The mode at 41~cm$^{-1}$ was observed by THz-transmission spectrocopy \cite{Kurumaji:2017a,Vasin:2024}.}
\begin{ruledtabular}
\begin{tabular}{cc|cc|c}
\multicolumn{2}{c|}{IR\cite{Reschke:2020}}          & \multicolumn{2}{c|}{Raman \cite{Stanislavchuk:2020}}   &  Mode \\
\multicolumn{2}{c|}{$\textbf{E}^\omega\parallel a$}   & \multicolumn{2}{c|}{$y(xz)\bar{y}$}  & assignment       \\
70\,K & 5\,K                  & 85\,K & 5\,K               & $E_{1}$(i) $i=1,\dots,12$  (calc.)\\
\hline
135   & 129   & -    & -     & 162        \\
-     & -     & 191  & 194   & 191      \\
218   & 214   & 216  & 213   & 223      \\
290   & 292   & -    & -     & 286      \\
-     & -     & -    & -     & 312      \\
333   & 335   & 327  & 333   & 351      \\
454   & 452   & 451  & 454   & 455      \\
473   & 471   & -    & -     & 473      \\
-     & -     & 481 & 487  & 481      \\
510   & 514   & 504  & 500   & 522      \\
561   & 559   & 565  & 575   & 577      \\
751   & 750   & -    & 748  & 750      \\
\hline
-     & -     & 264  & -  & $A_1$ at 263 (exp.) \\
-     & -     & 769  & 769   & $A_1$ at 770 (exp.) \\
\hline
-     & 41 \cite{Kurumaji:2017a,Vasin:2024} & - & - & $E_2$(1) at 41 (exp.) \\
-     & -     & 169  & 179   & not identified      \\
-     & -     & -     & 253   & electronic at 256 (calc.) \\
-     & 270   & -     & -     & $E_2$(5) at 268 (exp.) \\
-     & 426   & -     & -     & electronic at 426 (calc.)  \\
-     & 468   & -     & -     & $E_2$(9) at 469 (exp.) \\
-     & -     & 242  & 242   &  $B_1$(3) at 253 (calc.)       \\
\end{tabular}
\end{ruledtabular}
\end{table}

The comparison for the $E_1$ modes in Tab.~\ref{tab:xz_fmo} reveals deviation between experimental and theoretical eigenfrequencies for some of the  modes. For example,  the lowest-lying phonon mode $E_1(1)$, is predicted at 162~cm$^{-1}$, but clearly observed in the IR reflectivity at 129~cm$^{-1}$ with about 20\% deviation from the calculated value, clearly larger than any deviation in \cmo{}. In addition, resonant Raman scattering effects similar to \cmo{} are observed with respect to $A_1$ and $B_1$ modes already at RT, and magnetic symmetry breaking is evidenced by the appearance of several $E_2$ modes in the antiferromagnetic phase. Note that the mode at 41~cm$^{-1}$ was observed by THz-transmission spectroscopy \cite{Kurumaji:2017a,Vasin:2024} and identified as a Raman active $E_2$ mode by Wu \textit{et al}.~\cite{Wu:2023}. Again, one additional low-temperature mode is identified as an Fe$^{2+}$ multiplet state, and one mode visible at both reported temperatures remains unidentified.

Finally, the comparison for the $E_2$ modes is given in Tab.~\ref{tab:xy_fmo} and similarly to the lowest lying $E_1$ mode, mode $E_2(1)$ at 41~cm$^{-1}$ is observed at a significantly lower eigenfrequency in comparison to the expected one at 77~cm$^{-1}$, a deviation of about 45\%. The rest of the $E_2$ modes also deviate slightly stronger with the calculation and with the observation in \cmo, but the overall agreement is still good, including that mode $E_2(2)$ is not observable at a laser frequency of 532~nm, which was also used in Ref.~\onlinecite{Stanislavchuk:2020}. With respect to resonant Raman effects related to $A_1$ or $B_1$ modes, only one mode appearing above the antiferromagnetic transition was assigned to the $B_1(3)$ mode also visible in $y(xz)\bar{y}$ configuration. At low temperature, the $E_1$(1) mode indicates the magnetic symmetry breaking and three modes were assigned to calculated Fe$^{2+}$ multiplet states reported by Vasin \textit{et al}.\,\cite{Vasin:2024}.

To sum up, the conclusions drawn for \cmo{} regarding the mode assignment remain valid also for \fmo{}, but the lowest lying $E_1$ and $E_2$  (or $DE$) phonons experience a clear redshift in comparison with the \textit{ab initio} calculations. For the lowest lying mode, it was suggested that this shift is due to the hybridization of the mode with the closest lying magnon, resulting in a distinct chirality of $E_2(1)$ in applied magnetic fields \cite{Wu:2023}. Moreover, we may conclude that Raman scattering at different wavelengths may also reveal the elusive phonon modes in \fmo{}.

\begin{table}[t]
\caption{\label{tab:xy_fmo}
Comparison of experimental IR- and Raman-active  excitation frequencies (in cm$^{-1}$) in \fmo{} measured  above and below the magnetic phase transition in the Raman configuration $z(yx)\bar{z}$ \cite{Stanislavchuk:2020}, respectively. 
Mode assignment is made by comparison with calculated phonon eigenfrequencies for $E_2$ modes and by comparing to experimental values from other Raman measurement configurations and calculated Fe$^{2+}$ multiplet states reported by Vasin \textit{et al}.\,\cite{Vasin:2024}. The data for mode $E_2(1)$ is taken from \cite{Wu:2023}.}
\begin{ruledtabular}
\begin{tabular}{cc|c}
\multicolumn{2}{c|}{Raman}   & DFT +$U$ \\
\multicolumn{2}{c|}{$z(yx)\bar{z}$}  & AFM      \\
85\,K & 5\,K & $E_{2}$ $i=1,\dots,12$  (calc.)\\
 
\hline
46 \cite{Wu:2023}   & 41 \cite{Wu:2023}     & 77        \\
 -   & -   & 147      \\
  176  & 180    & 191      \\
211  & 205   & 215      \\
267  & 268   & 277         \\
328  & 328   & 334      \\
333  & 334   & 359      \\
448  & 448   & 453      \\
469  & 469   & 472      \\
513  & 513   & 482    \\
555  & 555   & 525      \\
-  & -   & 572      \\
737  & 746  & 751      \\
\hline
-    & 127    & $E_1$(1) at 129 (exp.) \\
-    & 158    & electronic at 165 (calc.)  \\
-    & 224    & electronic at 233 (calc.)  \\
 240    & -     & $B_1$(3) at 253 (calc.) \\
-    & 253    & electronic at 256 (calc.) \\
\end{tabular}
\end{ruledtabular}
\end{table}

\clearpage

\section{Additional IR  and Raman   spectra} \label{app:Ramanspectra}

\begin{figure}[h!]
    \centering
    \includegraphics[width=\linewidth]{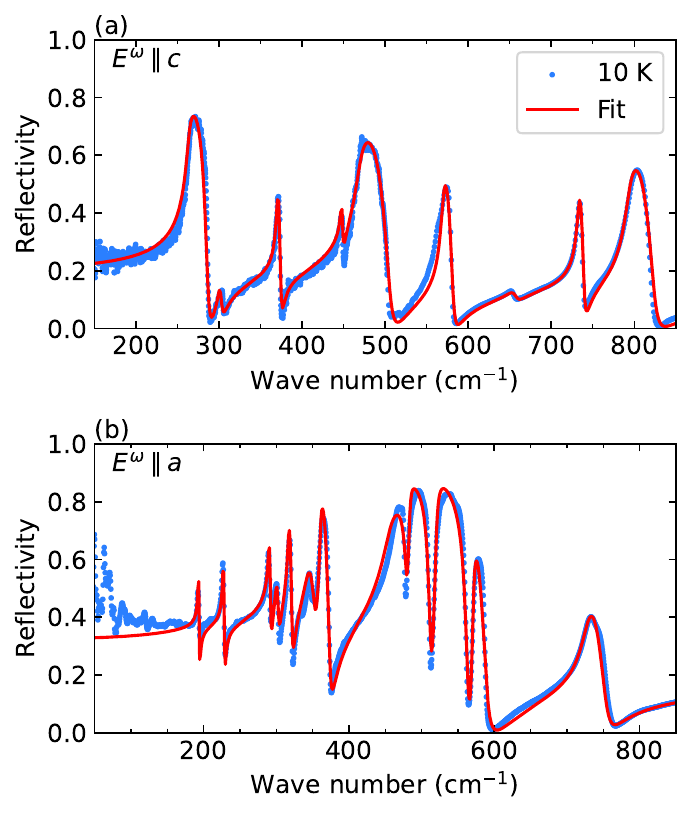}
    \caption{Reflectivity spectra in \cmo{} at 10~K and corresponding fit curves using Lorentzian oscillators for (a) $E^{\omega}\parallel c$ and (b) $E^{\omega}\parallel a$.}
    \label{fig:IR_fit}
\end{figure}

In Fig.~\ref{fig:IR_fit}(a) and (b) we show reflectivity spectra for \cmo{} at 10~K for light polarisation $E^{\omega}\parallel c$  and $E^{\omega}\parallel a$, respectively, together with fits using Lorentzian oscillators and the RefFIT software \cite{Kuzmenko:2005}. The obtained eigenfrequencies are listed in Tables \ref{tab:zz} and \ref{tab:xz}.

\begin{figure*}[t]
    \centering
    \includegraphics[width=\linewidth]{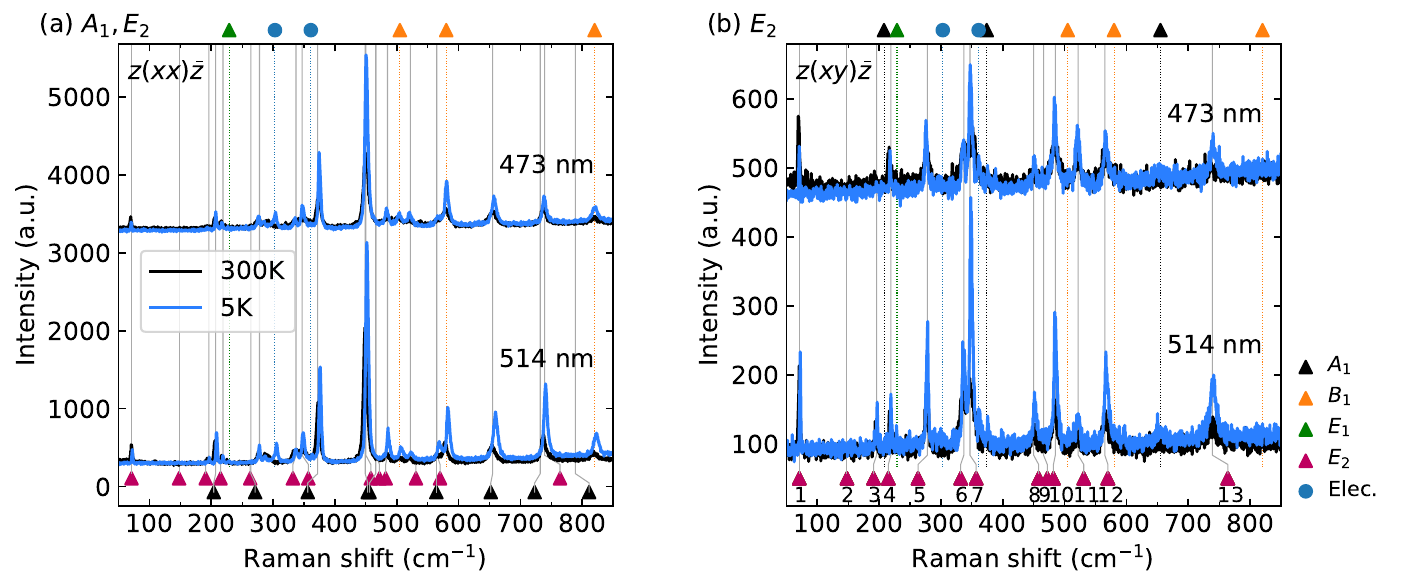}
\caption{Comparison of Raman spectra at laser wavelengths of 514~nm and  473~nm in scattering configurations (a) $z(xx)\bar{z}$  allowing for the observation of $E_2$ and $A_1$ modes, and (b)  $z(xy)\bar{z}$  corresponding to the irreducible $E_2$ representations of the paramagnetic phase. The calculated phonon eigenfrequencies expected for each configuration in the paramagnetic phase are indicated by upward triangles at the bottom of the individual panels. Symbols at the top of the panels indicate modes different from expected modes of the paramagnetic phase as described in the main text.}
    \label{fig:xx_xy_473_514}
\end{figure*}

In Fig.~\ref{fig:xx_xy_473_514} we show and compare the additional spectra for the Raman scattering configurations $z(xx)\bar{z}$ and $z(yx)\bar{z}$  for laser wavelengths of 514~nm and 473~nm measured in the paramagnetic phase at 300~K and at 5~K. The eigenfrequencies are listed in Tab.~\ref{tab:xyresonant} and were discussed with the spectra taken with wavelengths of 532~nm and 633~nm. 

\begin{figure*}[h!]
    \centering
    \includegraphics[width=\linewidth]{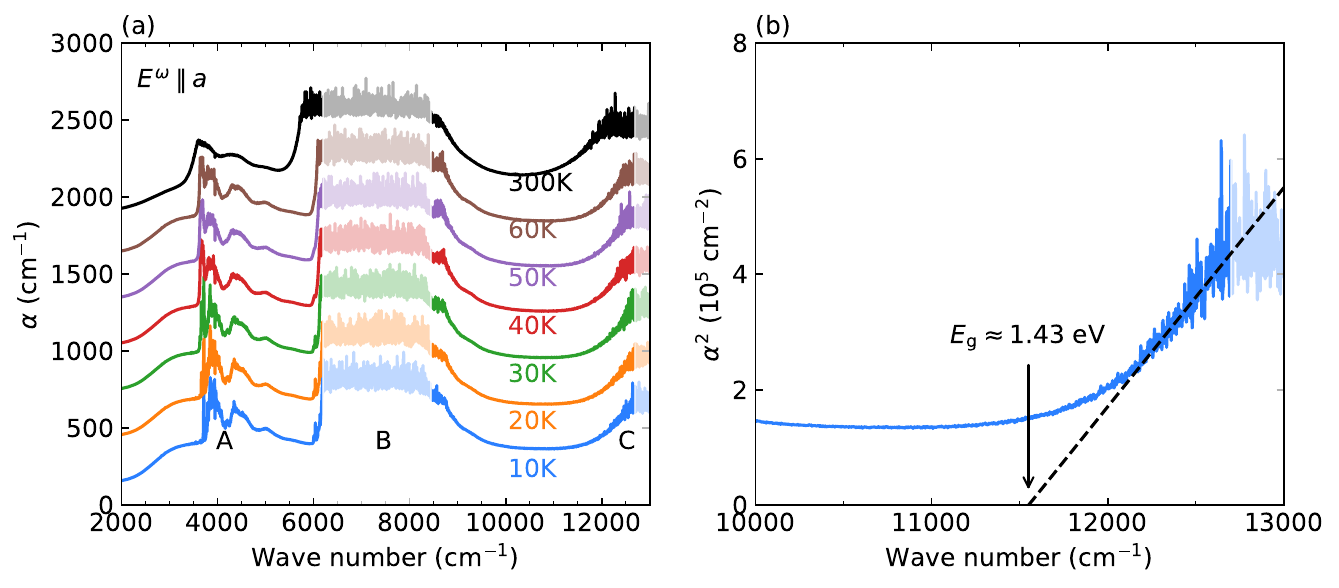}
    \caption{(a) Temperature dependent absorption spectra for $E^{\omega}\parallel a$ in \cmo{} revealing three excitation bands $A,B,C$ in the MIR/NIR frequency range. The semi-transparent ranges indicate bands of zero transmission through the sample. (b) Plot of $\alpha^2$ vs.~ wave number for the onset of band $C$ yields an estimate of 1.43~eV for the direct band gap.}
    \label{fig:absorption}
\end{figure*}

In Fig.~\ref{fig:absorption}(a), absorption coefficient spectra  for light polarization $E^{\omega}\parallel a$  in \cmo{} are shown in the mid-infrared (MIR) and near-infrared (NIR) frequency regime for several temperatures crossing the antiferromagnetic ordering transition at $T_\mathrm{N}=40$~K. The absorption coefficient was determined directly from the transmission coefficient $T$ via $\alpha=-1/d\ln{T}$ with the thickness $d$ of the sample.
We identify three main broad bands $A,B,C$, which are visible at all temperatures. The maxima of bands $B$ and $C$  could not be resolved due to the strong absorption. However, band $A$ develops a clear fine structure in the antiferromagnetic phase with a prominent narrow peak arising at 3719~cm$^{-1}$. For band $B$ only the flanks of the excitation peak can be resolved, but on the low-energy flank an emergence of a fine structure with a peak at 6038~cm$^{-1}$ is also observable. For band $C$ only the low-energy rise could be resolved before the sample becomes optically opaque for higher frequencies.

We interpret the origin of bands $A$ and $B$ in connection to excitations of the Co$^{2+}$ multiplet states for the tetrahedral and octahedral sites. In comparison with other compounds with Co$^{2+}$ in tetrahedral environment such as Co$_3$O$_4$ \cite{Mironova:1994} the origin of band $A$ is assigned to the $A$-site Co$^{2+}$ ions, while the assignment of band $B$ is not
clear at present. For example, in CoCr$_2$O$_4$  with only tetrahedrally coordinated Co$^{2+}$ ions an excitation band starting at about 6200~cm$^{-1}$ has been observed \cite{Kocsis:2018a}, while in several compounds with Co$^{2+}$ in octahedral environment similarly strong band appears in the same frequency region \cite{Ferguson:1963,Kant:2008}. Possibly, both Co sites contribute to band $B$. Moreover, in \fmo{} similar NIR features with fine structure have been reported \cite{Stanislavchuk:2020,Park:2021,Vasin:2024}, but the additional separated band $C$ was not observed. The origin of the fine structure is not easily determined, but vibrational and magnon sidebands are the usual candidates for such features \cite{Deisenhofer:2008,Kocsis:2018a,Park:2021,Vasin:2024}. We interpret band $C$ as the onset of the semiconducting direct band gap and roughly estimate a gap value of $E_g=1.43$~eV at 10~K from the plot shown in Fig.~\ref{fig:absorption}(b). This value is not too far from reported band structure calculations \cite{Park:2021}, which estimate a gap of about 1.55~eV for \cmo{}, and a somewhat lower gap value  and a merging and hybridzation of states forming bands $B$ and $C$ for \fmo{}. In any case, the used laser frequencies for the Raman experiments in both \cmo{} and \fmo{} are clearly above the band gap for both compounds and allow for resonant Raman effects.

\end{document}